\newcommand{\be}{\begin{equation}}
\newcommand{\ee}{\end{equation}}
\newcommand{\bea}{\begin{eqnarray}}
\newcommand{\eea}{\end{eqnarray}}
\newcommand{\nn}{\nonumber}\newcommand{\p}[1]{(\ref{#1})}
\newcommand{\lb}{\label}

\newcommand\q{\quad}
\newcommand\qq{\qquad}
\newcommand{\bt}{\bar{\theta}}
\newcommand{\pp}{{=\!\!\!|}}
\newcommand{\xp}{x^\pp}
\newcommand{\xm}{x^=}
\newcommand{\xpl}{x^\pp_L}
\newcommand{\xml}{x^=_L}
\newcommand{\xpr}{x^\pp_R}
\newcommand{\xmr}{x^=_R}
\newcommand{\Pp}{\partial_\pp}
\newcommand{\Pm}{\partial_=}

\newcommand{\tp}{\theta^+}
\newcommand{\btp}{\bar{\theta}^+}
\newcommand{\tm}{\theta^-}
\newcommand{\btm}{\bar{\theta}^{-}}

\newcommand{\Dp}{D_+}
\newcommand{\Dm}{D_-}

\newcommand{\bDp}{\bar{D}_+}
\newcommand{\bDm}{\bar{D}_-}
\newcommand{\tpo}{\theta^+_1}
\newcommand{\tpt}{\theta^+_2}
\newcommand{\tmo}{\theta^-_1}
\newcommand{\tmt}{\theta^-_2}
\newcommand{\Dpo}{D_+^1}
\newcommand{\Dmo}{D_-^1}
\newcommand{\Dpt}{D_+^2}
\newcommand{\Dmt}{D_-^2}

\newcommand{\ep}{\epsilon^+}
\newcommand{\emi}{\epsilon^-}
\newcommand{\bep}{\bar{\epsilon}^{\,+}}
\newcommand{\bem}{\bar{\epsilon}^{\,-}}

\newcommand{\ept}{\epsilon^+_2}
\newcommand{\emt}{\epsilon^-_2}

\documentclass[11pt]{article}

\begin{document}
\begin{titlepage}
\begin{flushright}
ITP-UH-38/00 \\
JINR-E2-2000-290\\
hep-th/0012199\\
December 2000
\end{flushright}
\vskip 0.6truecm
\begin{center}{\Large\bf
Partial spontaneous breaking \\[.3cm]
of two-dimensional supersymmetry}
\end{center}
 \vskip 0.6truecm
\centerline{ E. Ivanov${}^{\,a,1}$, S. Krivonos${}^{\,a,2}$,   
O. Lechtenfeld${}^{\,b,3}$, B. Zupnik${}^{\,a,4}$  }  
\vskip 0.6truecm
\centerline{$^a${\it Bogoliubov Laboratory of
Theoretical Physics, JINR,}}
\centerline{\it 141 980 Dubna, Moscow region,
Russian Federation}

\vspace{0.2cm}
\centerline{$^b$ {\it Institut f\"ur Theoretische Physik, Universit\"at
Hannover,}}
\centerline{\it  Appelstra\ss{}e 2, 30167, Hannover, Germany}

\vspace{0.2cm}
\vskip 0.6truecm  \nopagebreak

\begin{abstract}
\noindent We construct low-energy Goldstone 
superfield actions describing various patterns of the partial spontaneous 
breakdown of two-dimensional $N=(1,1)$, $N=(2,0)$ and $N=(2,2)$
supersymmetries, with the main focus on the last case. These
nonlinear actions admit a representation in the superspace of the unbroken 
supersymmetry as well as in a superspace of the full supersymmetry. The
natural setup for implementing  the partial breaking in a
self-consistent way is provided by the appropriate central extensions of
$D=2$ supersymmetries, with the central charges generating shift symmetries
on the Goldstone superfields. The Goldstone superfield actions can be
interpreted as manifestly world-sheet supersymmetric actions in the 
static gauge of some superstrings and D1-branes in $D=3$ and $D=4$
Minkowski spaces. As an essentially new example, we elaborate on the action
representing the $1/4$ partial breaking pattern $N=(2,2) \rightarrow
N=(1,0)$. 
\end{abstract}

\vfill

\noindent{\it E-Mail:}\\
{\it 1) eivanov@thsun1.jinr.ru}\\
{\it 2) krivonos@thsun1.jinr.ru}\\
{\it 3) lechtenf@itp.uni-hannover.de}\\
{\it 4) zupnik@thsun1.jinr.ru}
\newpage

\end{titlepage}

\renewcommand{\thefootnote}{\arabic{footnote}}
\setcounter{footnote}0
\setcounter{equation}0
\section{Introduction}
Supersymmetric models in two dimensions, in particular $N{=}(2,2)$
ones, were a subject of numerous studies (see, e.g.,\cite{GHR}-\cite{MP}). 
The generic source of interest in such models is their tight relation to
strings, integrable systems  and $D=2$ conformal theory. In particular, 
$D=2$ conformal field theories with $N=(2,2)$ supersymmetry are considered
as candidate vacua for perturbative string theory. More specifically,
$N=(2,2)$ models, being invariant under the simplest extended supersymmetry
in $D=2$ and allowing for concise off-shell superfield formulations
\cite{GHR}-\cite{RV}, provide the proper laboratory for analyzing the 
characteristic features they share with more complicated higher-dimensional
superfield theories.

One of such generic features is the phenomenon of spontaneous partial
breaking of  global supersymmetry (PBGS) \cite{BW}-\cite{W}. The
first self-consistent example of the Goldstone-fermion $D=2$ model with the
spontaneous partial breaking of global  supersymmetry was constructed
in ref. \cite{HP}. There, the partial breaking $N=(2,2)\;\rightarrow \;
N=(2,0)$
was triggered by a topologically non-trivial BPS classical solution preserving
one half of the original supersymmetry, the one generated by two left
supercharges only. The full $N=(2,2)$ supersymmetry algebra was found to
include two central charges (also spontaneously broken), and the
resulting invariant action proved to be the static gauge form of
the  Green-Schwarz action of $N=1, D=4$ superstring. The construction of
ref.\cite{HP} essentially exploited the methods of nonlinear realizations of
global supersymmetry. Recently, these methods were 
applied to study various PBGS options in the models with
$D{=}4,~N{=}2$ \cite{BG2}-\cite{GPR}, $D{=}4,~N{=}4$ (or $D{=}10,
~N=1$) \cite{BIK} and $D{=}3,~N{=}2$ \cite{IK} supersymmetries, and to
reveal their relationships with branes. The interplay between the
PBGS description of branes and the one based on the superembedding
approach (\cite{Dima} and references therein) was studied in
\cite{swed}, \cite{PST}. 

There still remain some problems with the treatment of the PBGS phenomenon
which may be clarified by further studying it within
$D=2$ models. In particular, this regards the linear realizations
of  PBGS and their relation to the universal nonlinear
realizations  description. Also, besides the PBGS option leading to the
$N=1, D=4$ superstring along the line of ref. \cite{HP}, there exist others
which have not been properly investigated until now. 

The basic aim of the present paper is to fill this gap. We construct the
manifestly worldvolume supersymmetric Goldstone superfield actions for the
PBGS patterns  $N=(2,2) \rightarrow N=(1,1)$ and $N=(2,2) \rightarrow
N=(1,0)$ and give them an interpretation as the static-gauge actions of
some superstrings and D1-branes. We also reproduce, in a new setting, the
$N=(2,2) \rightarrow N=(2,0)$ model of ref. \cite{HP}. As a by-product, we
find some interesting examples of PBGS in the cases of $N=(1,1)$ and
$N=(2,0)$ supersymmetries. All the Goldstone superfield  actions are written 
in a two-fold way: as integrals over superspaces of unbroken supersymmetry
and as integrals over superspaces of the full spontaneously
broken supersymmetry, which is manifest in the latter formulation. This
allows us to unveil the relationship between linear
and nonlinear realizations of the partial breaking patterns considered. 
For self-consistency, in all non-trivial cases it proves necessary to
proceed from the properly central-charge extended $D=2$ supersymmetry. 
These central charges generate shift symmetries of the Goldstone
superfields and play a crucial role both in the implementation of the 
partial breaking and in the superbrane interpretation of the Goldstone 
superfield actions.      

In our study, we systematically make use of the general relationship
between linear and nonlinear realizations of supersymmetry \cite{IKa}.
It was already applied to the PBGS case in a recent paper \cite{DIK}. 

\vfill\eject

\setcounter{equation}0
\section{\lb{A} ABC of N=(2,2), D=2 superspace}
The aim of this Section is to adduce the necessary information 
about the superspaces of $N=(2,2)$, $D=2$ supersymmetry which is the 
basic object of our study in the present paper. We also focus on the
structure of central-charge extensions of this supersymmetry.   

The full $D{=}2,~N=(2,2)$ superspace $z = (z_l, z_r)$ consists of two
light-cone sectors, the left (2,0) and right (0,2) ones $z_l$ and $z_r$,
parametrized, respectively, by the coordinates 
\be
z_l=(\xp ,\tp ,\btp )~, \qquad z_r=(\xm ,\tm ,\btm )  \lb{A2}
\ee
with the $SO(1,1)$ weights $(\pm2,\pm1,\pm1)$. 
Sometimes it is convenient to parametrize $N=(2,2)$ superspace
by four real Grassmann coordinates $\theta_1^\pm~, \,\theta_2^\pm$,
\bea
&&\theta_1^\pm={1\over\sqrt{2}}(\theta^\pm +\bt^\pm )~,\qquad
\theta_2^\pm={i\over\sqrt{2}}(\bt^\pm -\theta^\pm)~, \nn \\
&& \theta^\pm = {1\over\sqrt{2}}(\theta^\pm_1 +i \theta^\pm_2)~, \qquad
\bar\theta^\pm = {1\over\sqrt{2}}(\theta^\pm_1 -i \theta^\pm_2)~. \lb{A4}
\eea

The algebra of spinor derivatives in the (2,0) and (0,2) sectors  has
the following form:
\bea
&& \{\Dp ,\bDp\}=2i\Pp\equiv 2P_\pp~,\qq \Dp\Dp=0~,\qq \bDp\bDp=0~, \lb{A5}\\
&& \{\Dm ,\bDm\}=2i\Pm\equiv 2P_=~,\qq \Dm\Dm=0~,\qq \bDm\bDm=0~.\lb{A6}
\eea
The crossing relations admit four real $SO(1,1)$ singlet central charges
\bea
&& \{\Dp,\bDm\}=2(Z_1 +iZ_2)~,\qq \{\bDp,\Dm\}=2(Z_1 -iZ_2)~,
\nn\\
&& \{\Dp,\Dm\}=2(Z_3 +iZ_4)~,\qq \{\bDp,\bDm\}=2(Z_3 -iZ_4)~.\lb{A8}
\eea
These spinor covariant derivatives can be chosen in the following explicit
form:
\bea 
&&D_+ = \frac{\partial}{\partial \theta^+} -i \bar\theta^+ \Pp 
- \bar\theta^- (Z_1 +iZ_2) + \theta^- (Z_3 +iZ_4)~, \nn \\ 
&&\bar D_+ = -\frac{\partial}{\partial \bar\theta^+} +i \theta^+ \Pp 
- \bar\theta^- (Z_3 -iZ_4) + \theta^- (Z_1 -iZ_2)~, \nn \\
&&D_- = \frac{\partial}{\partial \theta^-} -i \bar\theta^- \Pm 
- \bar\theta^+ (Z_1 -iZ_2) + \theta^+ (Z_3 +iZ_4)~, \nn \\
&&\bar D_- = -\frac{\partial}{\partial \bar\theta^-} +i \theta^- \Pm 
- \bar\theta^+ (Z_3 -iZ_4) + \theta^+ (Z_1 +iZ_2)~. \lb{explicD}
\eea
Note that this definition is not unique in view of possibility to  
perform similarity transformations 
\be
(D, \bar D)  \quad \Rightarrow \quad (\tilde{D}, \tilde{\bar D}) = 
e^{-A}(D, \bar D)e^{A}~, \lb{simil}
\ee
where the operator $A$ is a linear combination of central charges with 
the coefficients bilinear in Grassmann coordinates (all differential 
operators, e.g., generators of supersymmetry, and $N=(2,2)$ superfields 
should be simultaneously transformed). For instance, 
making such a transformation with 
\be
A = -\theta^+\theta^-(Z_3 +iZ_4) - \bar\theta^+\bar\theta^-(Z_3 -iZ_4)
+ \bar\theta^+\theta^-(Z_1 -iZ_2)+ \theta^+\bar\theta^-(Z_1 +iZ_2)~,
\ee
one can entirely remove the central-charge terms from $D_+, \bar D_+$ 
at cost of appearance of the factor 2 in front of such terms in 
$D_-, \bar D_-$. In this particular case $A = - A^\dagger$, so the 
transformation preserves the simple properties of covariant derivatives
under complex conjugation. In other cases the conjugation rules 
become more complicated.   

For the spinor derivatives in the real parametrization \p{A4} 
\be
D^1_\pm=
{1\over\sqrt{2}}
(D_\pm -\bar{D}_\pm)~, \qquad D^2_\pm=
{i\over\sqrt{2}}
(D_\pm +\bar{D}_\pm)~,
\lb{A9b}
\ee
eqs. \p{A5} - \p{A8} imply the following relations:
\bea
&& \Dpo\Dpo=\Dpt\Dpt= -i\Pp~,\qq \Dmo\Dmo=\Dmt\Dmt= -i\Pm~,\lb{A10}\\
&& \{\Dpo,\Dpt\}=0~,\qq \{\Dmo,\Dmt\}=0~.\lb{A11} \\
&& \{\Dpo,\Dmo\}= 2(Z_3-Z_1)~,\quad\{\Dpt,\Dmt\}= -2(Z_1+Z_3)~,\lb{A12}\\
&& \{\Dpo,\Dmt\}= -2(Z_2+Z_4)~,\quad\{\Dpt,\Dmo\}=2(Z_2-Z_4)~.\lb{A13}
\eea
The explicit form of these derivatives can be easily found from \p{explicD}
and \p{A4}, for instance
\be
\Dpo = \frac{\partial}{\partial \theta^+_1} -i \theta^+_1 \Pp 
- \theta^-_1 (Z_1 - Z_3) -\theta^-_2 (Z_2 +Z_4)~. \lb{D1}
\ee

We shall use the following notation for the infinitesimal  $N=(2,2)$
supersymmetry transformation of some $N=(2,2)$ superfield $\phi(z)$:
\bea
&&\delta \phi =(\epsilon Q)\phi ,\nn\\
&&(\epsilon Q)=\ep Q_+ +\bep\bar{Q}_+ +\emi Q_- +\bem\bar{Q}_- =\ep_k Q_+^k
+\emi_k Q_-^k=-(\epsilon Q)^\dagger \lb{A23}
\eea
and the corresponding finite transformation is
\be
\phi{}'(z) \equiv \phi (\epsilon,z)=\mbox{exp}\{\epsilon
Q\}\phi(z)~.\lb{A23b}
\ee
Here $k = 1,2$ and 
$$
Q^1_\pm = {1\over \sqrt{2}}(Q_\pm + \bar Q_\pm)~, \qquad 
Q^2_\pm = {i\over \sqrt{2}}(Q_\pm - \bar Q_\pm)~. 
$$
With this convention, the algebra of $N=(2,2)$ supersymmetry has the form: 
\bea
&&\{Q_+,\bar{Q}_+\}=2P_\pp~,\q \{Q_-,\bar{Q}_-\}=2P_=~,\nn\\
&&\{Q_+,\bar{Q}_-\}=2(Z_1 +iZ_2)~,\q \{Q_+,Q_-\}=2(Z_3 +iZ_4)~.
\lb{A23c}
\eea
The algebra of generators in the real parametrization \p{A4}, i.e. 
that of $Q^1_\pm, Q^2_\pm$, can be easily read off from these 
relations, e.g.,    
\be
Q^1_+Q^1_+ = i\Pp~,\quad Q^1_-Q^1_- = i\Pm~, \quad \{Q^1_+, Q^1_-\} =
2(Z_1+ Z_3)~. \lb{q1alg2}
\ee

In what follows we shall deal with the $N=(2,2)$
superspace  without special coordinates for the central charges. 
\footnote{Adding two such coordinates would actually mean passing to
$N=1, D=4$ superspace: $N=(2,2)$ superalgebra with one complex
central charge in the crossing anticommutators (it can be $Z_1\pm iZ_2$
or $Z_3\pm iZ_4$, these choices are equivalent up to the twist $D_-
\leftrightarrow \bar
D_-$) is isomorphic
to the standard $N=1,~D=4$ Poincar\'e  superalgebra (total of 4 bosonic
generators). On the other hand, the situation with {\it two} independent
complex  central charges in \p{A8}
(total of 6 bosonic generators) corresponds to some non-trivial truncation of
the extension of $N=1,~D=4$ Poincar\'e superalgebra by complex tensorial
central charges \cite{FeP}-\cite{GGHT}. The full such extension (10
bosonic generators) in the
$D=2$ language would correspond to inserting two independent complex
charges with the appropriate $SO(1,1)$ weights also  into the r.h.s. 
of $D_{\pm}D_{\pm}$. Though such further modification of $N=(2,2)$
superalgebra can certainly have implications in the PBGS 
and brane contexts (see, e.g., \cite{DIK} for the case of superparticle),
for simplicity we shall not dwell on this possibility here.}
These charges will show up only as generators of some shifting isometries
realized on fields, i.e. as internal symmetry generators. Thus we shall
basically use (in particular, in the rest of this Section) the spinor
derivatives \p{explicD}, \p{D1} with the central charge 
terms neglected. The corresponding superalgebra and superspace reveal the
automorphism $U(1)_+\times U(1)_-$ symmetry (or $R$-symmetry) realized as
independent phase transformations of the left and right pairs of the spinor
coordinates 
\be
(\tp,\btp)~,\qquad
(\tm,\btm)~. \lb{A15} 
\ee
Actually, this symmetry is explicitly broken, at least down 
to the diagonal $U(1)$, in most $N=(2,2)$
models.\footnote{The extended superalgebra \p{A8} is still covariant
under these automorphisms, provided one ascribes appropriate
transformation properties to the complex central charges.}
The integration measure in the full $N=(2,2)$ superspace is defined by
\be
d^6z= \overline{(d^6z)} = 
d^2x\Dm\Dp\bDp\bDm=d^2x\Dpo\Dmo\Dpt\Dmt~.\lb{A16b}
\ee

In both light-cone sectors of $N=(2,2)$ superspace there exist
mutually conjugated left and right chiral (chiral and antichiral) 
subspaces. E.g., in the (2,0) sector 
the corresponding coordinate sets are as follows:
\bea
\zeta_L^{(+)} &=& (\xpl,\tp)~,\qquad \xpl=\xp-i\tp\btp=\xp-\tpo\tpt~,
\lb{A17} \\
\zeta_R^{(+)} &=& (\xpr,\btp)~,\qquad \xpr =\xp+i\tp\btp=\xp+\tpo\tpt
\lb{A18}~.
\eea
The chiral and antichiral coordinates in the (0,2) sector 
$\zeta_L^{(-)}= (\xml,\tm)$ and $\zeta_R^{(-)}= (\xmr,\tm)$ can be
introduced analogously. The integration measures in the chiral and
antichiral $N=(2,2)$ superspaces $\zeta_L = (\zeta_L^{(+)}), 
\zeta_L^{(-)})$ and $\zeta_R = (\zeta_R^{(+)}, \zeta_R^{(-)})$ 
are defined as 
\be
d^4\zeta_L = d^2\zeta_L^{(-)} d^2\zeta_L^{(+)} =  d\xpl d\xml\Dm\Dp~,\;\;
d^4\zeta_R = d^2 \zeta_R^{(+)}d^2 \zeta_R^{(-)} = d\xpr
d\xmr\bDp\bDm~.\lb{A19}
\ee

The chiral $N=(2,2)$ superfields are defined by the following
constraints 
\be
\bDp\phi=\bDm\phi=0~, \lb{A20}
\ee
while the twisted-chiral ones \cite{GHR,IK0} by 
\be
\Dp \lambda=\bDm \lambda =0~.\lb{A21}
\ee
The defining constraints for the antichiral and twisted anti-chiral
superfields follow from \p{A20}, \p{A21} by complex conjugation. Note 
that in the case of non-vanishing central charges the constraints \p{A20}, 
\p{A21} require, respectively, 
\be
(Z_3 -iZ_4)\phi = 0 \quad \mbox{and} \quad (Z_3 +iZ_4)\bar\phi = 0~, 
\lb{chirint}
\ee
or 
\be
(Z_1 +iZ_2)\lambda = 0  \quad \mbox{and} \quad (Z_1 -iZ_2)\bar\lambda = 0~.
\lb{twistint}
\ee 
as their integrability conditions. If these conditions are satisfied, the 
remaining central-charge terms can be removed from the spinor covariant
derivatives in \p{A20} or \p{A21} by a proper transformation 
like \p{simil}. As the result, in such a frame chirality or twisted 
chirality become manifest despite the presence of central charges 
in the $N=(2,2)$ superalgebra. For instance, if some superfield $\phi$
satisfies the conditions \p{A20} with $\bar D_\pm $ defined by \p{explicD},
the superfield 
$$
\tilde \phi = e^{-A}\phi~, \quad A = \bar\theta^+\theta^-(Z_1 -i Z_2) - 
\theta^+ \bar\theta^-(Z_1 +i Z_2)
$$
satisfies the standard chirality conditions with ``short'' covariant
derivatives (containing no central-charge terms), and so it lives on 
the chiral superspace $\zeta_L  = (\zeta^{(+)}_L, \zeta^{(-)}_L)$.
       
In what follows we shall need some facts about $N=(1,1)$, $N=(2,0)$
and $N=(1,0)$ superspaces as  subspaces of  the $N=(2,2)$ one. All 
the covariant spinor derivatives below are assumed to be ``short'', i.e. 
containing no central-charge terms.

The $N=(1,1)$ superspace is parametrized by the real spinor coordinates
$\tpo,\tmo$ related to the $N=(2,2)$ ones according to \p{A4}. 
The simplest free $N=(1,1)$ scalar multiplet model is described by the
unconstrained real scalar superfield $\pi(x,\tpo,\tmo)$ with the action 
\be
S={1\over2i}\int d^4z \,\Dpo\pi\Dmo\pi \lb{A34}
\ee
where 
\be
d^4 z = \overline{(d^4z)} \equiv d^2x\,i\Dpo\Dmo \lb{11meas}
\ee
is the $N=(1,1)$ superspace integration measure. 

The $N=(2,0)$ and $N=(0,2)$ subspaces (not to be confused with the
$(2,0)$ and $(0,2)$ light-cone sectors) are spanned by the 
following coordinate sets
\be 
z^{(+)} = (\xm, z_l) = (\xm, \xp, \tp, \btp)~, \quad z^{(-)}= (\xp, z_r) =
(\xp, \xm, \tm, \btm)~.
\ee
The appropriate integration measures have non-trivial $SO(1,1)$ weights
($\mp 2$):
\be
d^4z^{(+)} = d^2x \Dp \bDp = i d^2x D^2_+D^1_+~, \quad d^4z^{(-)} = d^2x
\Dm \bDm = i d^2x D^2_-D^1_-~.   
\ee
The simplest  off-shell representations are comprised by the $N=(2,0)$
and $N=(0,2)$ chiral superfields $\varphi(\xpl, \xm, \tp)$ and  
$\omega(\xp, \xml, \tm)$ with the following free actions 
\be
S_0(\varphi)\sim {1\over 2i}\int d^4z^{(+)}\,\bar{\varphi}\Pm\,\varphi~, 
\quad S_0(\omega)\sim {1\over 2i}\int d^4z^{(-)}\,\bar{\omega}\Pp\,\omega~.
\lb{A37}
\ee

At last, the free action of an unconstrained scalar $N=(1,0)$ superfield
$v(\xp, \xm, \tpo)$ has the following form:
\be
{1\over 2}\int d^3z^{(+)} \,\Dpo v \Pm v~, \quad d^3z^{(+)} \equiv d^2x
\,i\Dpo
~.\lb{A38}
\ee
Its $N=(0,1)$ counterpart is constructed in a similar and evident way.

\setcounter{equation}0
\section{Toy examples}
To clarify the basic features of our approach we start with
the simplest examples of partial breaking of the $D=2$ global supersymmetry
with two supercharges. In $D=2$ there exist two
different patterns for such breaking, viz.  $N=(1,1)\rightarrow N=(1,0)$
and $N=(2,0)\rightarrow N=(1,0)$.

\subsection{$N=(1,1) \rightarrow N=(1,0)$}

To describe this PBGS option, we should construct $N=(1,0)$
superfield action possessing one additional spontaneously broken 
$N=(0,1)$ supersymmetry. The $N=(1,0)$ superfield formulation 
is preferable because only the $N=(1,0)$ supersymmetry is supposed to
remain unbroken and so manifest. As usual, the partial breaking implies the
presence of the Goldstone fermion among the component fields of our theory.
The simplest possibility is to start with a real bosonic scalar  $N=(1,0)$
superfield $u(\xp,\xm,\tp)$  and to define the real fermionic superfield
$\xi^-(\xp,\xm,\tp)$ \footnote{The real $N=(1,1)$ superspace coordinates
$\theta^\pm$ used in this Subsection should not be confused with the
complex $N=(2,2)$ superspace ones.} 
% $\xi_+(\xp,\xm,\tp)$
\be
\xi^- \equiv i D_+ u \;, \lb{defGF} 
\ee
with 
$$
D_+ = 
\frac{\partial}{\partial \theta^+} - i \theta^+ \Pp~, \;\; D_+D_+
= -i\Pp~,   
$$
whose first component is assumed to be the Goldstone fermion. Since 
the crucial characteristic feature of the latter is the pure
shift in the transformation rule under spontaneously broken 
$N=(0,1)$ supersymmetry,
$$
\delta \xi^- = \epsilon^- + \ldots~, 
$$   
the appropriate transformation of $u$ should also contain an
inhomogeneous term
$$
\delta u = i\epsilon^-\theta^+ + \ldots~.
$$
Here, $\epsilon^-$ is the transformation parameter 
 
In order to have a linear off-shell realization of this extra 
$N=(0,1)$ supersymmetry, we have to add one more fermionic superfield
$\eta^+(\xp,\xm,\tp )$. It is easy to find that the  
following transformation laws of $u$ and $\eta$ just constitute the
desired $N=(0,1)$ supersymmetry algebra
\bea\label{realiz1}
&&\delta u=i\epsilon^-\theta^+ +i \epsilon^-\eta^+\; \Rightarrow \quad
\delta\xi^-=\epsilon^- + \epsilon^-D_+\eta^+ \: , \nn \\
&&  \delta\eta^+= -\epsilon^-\partial_= u \;.
\eea
Together with the manifest $N=(1,0)$ supersymmetry it forms the full 
$N=(1,1)$, $D=2$  supersymmetry. 

It is easy to check that the closure of the manifest $N=(1,0)$ and hidden
$N=(0,1)$ supersymmetries on the superfield $u$ yields a constant shift
of $u$. So in the present case we are facing the central-charge
extended $N=(1,1), D=2$ supersymmetry algebra:
\be
\left\{ \hat{Q}_+,\hat{Q}_-\right\} = 2Z~, \;\; \hat{Q}_+\hat{Q}_+ = P_\pp 
= i\Pp~, \;\; 
\hat{Q}_-\hat{Q}_- = P_= = i\Pm\;, \lb{11Z}
\ee
where $Z$ acts as a pure translation of $u$:
\be
\delta_z u = -2i\,a [Z, u] = a~, \quad a^\dagger = a~,
\lb{Ztran}
\ee   
and ``hat'' was introduced to distinguish this algebra from the one without 
central charge.

It is instructive to see how this $N=(1,0)$ superfield representation could
equivalently be deduced from the $N=(1,1)$ superspace formalism.

To this end, let us consider a scalar real $N=(1,1)$ superfield
$\Phi(x,\theta^+,\theta^-)$ ($[\Phi] = -1$), 
\be
\Phi(x,\theta^+,\theta^-)=u(x,\theta^+)+i\theta^-\eta^+(x,\theta^+)~,\lb{1decom}
\ee
such that it possesses non-trivial transformation properties under the
central charge $Z$ which generates pure shifts of $\Phi$ (cf.
\p{Ztran}):
\be
\delta_z \Phi = -2i\,a[Z, \Phi] = a \;\; \Rightarrow \;\; 
Z \Phi = {i\over 2}~. \lb{PhiZ}
\ee 
For the moment, the coefficients in the $\theta^-$ expansion \p{1decom} of
$\Phi$ are arbitrary and not to be identified with the previous
$N=(1,0)$ superfields. The numerical coefficient in the definition of $Z$ 
in \p{PhiZ} was chosen for further convenience: it is defined up 
to an arbitrary rescaling of $\Phi$. 

The generators of the central-charge extended $N=(1,1)$ superalgebra
\p{11Z} can be chosen so as 
\bea
&& \hat{Q}_+ = \frac{\partial}{\partial \theta^+} + i \theta^+ \Pp~ = Q_+,
\nn
\\ 
&& \hat{Q}_- = \frac{\partial}{\partial \theta^-} + i \theta^- \Pm 
+ 2\theta^+ Z~ = Q_- + 2 \theta^+ Z. \label{QQgen} 
\eea
This choice is convenient in that $\Phi$ has the standard
superfield transformation law under $N=(1,0)$ supersymmetry and, hence, 
the coefficients in its $\theta^-$ expansion ($u$ and $\eta^+$) are 
automatically standard $N=(1,0)$ superfields \footnote{An equivalent 
choice of the generators, with $Z$ appearing in both of them, can be 
achieved by means of the appropriate transformation of the kind
\p{simil}.}. At the same time, $\Phi$, with taking into account \p{PhiZ},
inhomogeneously transforms under the second supersymmetry:
\be
\delta \Phi = \epsilon^-\hat{Q}_- \Phi = i\epsilon^- \theta^+ 
+ \epsilon^-Q_- \Phi~. \lb{transPhi}
\ee
It is straightforward to see that for the $N=(1,0)$ components of $\Phi$ 
\p{1decom} this law produces just the transformation laws \p{realiz1}. 
Hence, these components can be identified with the $N=(1,0)$ superfields 
introduced earlier.  

To construct invariants, we need to define the $N=(1,1)$ spinor covariant 
derivatives for this special case. These objects, anticommuting with the 
generators \p{QQgen}, are easily seen to be as follows
\bea
&& \hat{D}_+ = 
\frac{\partial}{\partial \theta^+} - i \theta^+ \Pp - 2\theta^-Z \equiv 
D_+ - 2\theta^-Z    
~, \nn \\ 
&& \hat{D}_- = \frac{\partial}{\partial \theta^-} - i \theta^- \Pm = 
D_-~, \quad D_+D_+ = -i\Pp~, \; D_-D_- = -i\Pm~.   
\label{DDgen} 
\eea   
So, for constructing $N=(1,1)$ invariants, we have two covariant
quantities:
\bea
&&\hat{D}_-\Phi = D_-\Phi = i (\eta^+ - \theta^- \Pm u)~, \nn \\ 
&&\hat{D}_+\Phi = D_+\Phi -i\theta^- = D_+u -i\theta^-(1
+D_+\eta^+)~.\lb{hatD}
\eea 
Then, like in the case without central charges, eq. \p{A34}, 
the simplest invariant action is as follows 
\be
S_\Phi = {1\over 2if^2} \int d^4z\,\hat{D}_+\Phi\,D_-\Phi = 
{1\over 2if^2} \int d^4z\,D_+\Phi\,D_-\Phi - {1\over 2f^2}\int d^4z\, \Phi 
\equiv S_1 - S_2~, \lb{act}
\ee
where we made use of \p{hatD} and integrated by parts with
respect to $D_-$ ($f$ is a normalization factor of dimension -1).
It is easy to see that $S_2$ is
invariant on its own because the
inhomogeneous term in the transformation law \p{transPhi} of $\Phi$ does
not contribute by the definition of the $N=(1,1)$ integration measure
\p{11meas}. So, $S_1$ is also invariant (up to a shift of the Lagrangian 
density by a full derivative). In terms of the $N=(1,0)$ superfields $u,
\eta^+$ these invariants read   
\bea
S_1 &=& {1\over 2f^2}\int d^3z^{(+)}(D_+u\Pm u - i\eta^+D_+\eta^+) \nn \\
&=& {i\over 2f^2}\int d^2x D_+(D_+u\Pm u - i\eta^+D_+\eta^+),
\lb{action1} \\
S_2 &=& {i\over 2f^2}\int d^3z^{(+)} \eta^+ = 
-{1\over 2f^2}\int d^2x D_+ \eta^+~,  \lb{action2}
\eea
where the $N=(1,0)$ superspace integration measure is defined in \p{A38}. 
We stress that only the combination \p{act} of these invariants is a genuine 
invariant of the central-charge extended $N=(1,1)$ supersymmetry; each of 
them individually is invariant up to the surface terms. This is analogous 
to what happens for WZNW or Chern-Simons actions.   

Before going further, let us summarize the above discussion. Starting
from the scalar $N=(1,0)$ multiplet $u$ and requiring its fermionic
component to be the Goldstone fermion corresponding to spontaneous partial 
breaking of $N=(1,1)$ supersymmetry down to
$N=(1,0)$, we uniquely restored the $N=(1,1)$ supermultiplet to which $u$
should belong. It proved to be a supermultiplet of the central-charge
extended $N=(1,1)$ superalgebra \p{11Z} and it is naturally accommodated 
by a scalar $N=(1,1)$ superfield $\Phi$ for which the central charge $Z$ 
generates pure shifts. The invariant action for this system is 
a combination of two independent invariants, $S_1$ and $S_2$. 
Thus what we have constructed can
be called a superfield model of {\it linear} realization of the partial
spontaneous breaking $N = (1,1)\;\rightarrow\; N=(1,0)$. 

Let us dwell on some peculiar features of this toy model which will show up 
as well in other $D=2$ PBGS examples we consider in this paper. 

As we saw, the presence of an inhomogeneously transforming 
fermionic Goldstone component in the scalar supermultiplet of unbroken 
supersymmetry ($N=(1,0)$) inevitably implies the appearance of
the central charge generator in the algebra of full spontaneously broken
supersymmetry ($N=(1,1)$ ). The symmetry generated by the central charge is
also spontaneously broken,  
the physical bosonic field of the scalar supermultiplet being the 
appropriate inhomogeneously transforming Goldstone boson ($N=(1,0)$
superpartner of Goldstino).

We may reverse the argument by starting from the central-charge 
extended $N=(1,1)$ superalgebra \p{11Z} and introducing the scalar 
superfield $\Phi(z)$ \p{1decom} which is shifted by a constant under the
action of the central charge generator (eq. \p{PhiZ}). So it can be
regarded as the Goldstone superfield associated with this spontaneously
broken generator. Its supersymmetry transformations are uniquely defined by
\p{QQgen}, \p{transPhi}. One half of these transformations, the $N=(0,1)$
ones, contain an inhomogeneous shift $i\epsilon^-\theta^+$, which implies
that $iD_+\Phi$ is shifted by $\epsilon^-$. Thus $iD_+\Phi$ is the Goldstone
fermionic superfield whose presence is tantamount to the spontaneous
breakdown of the half of $N=(1,1)$ supersymmetry. In other words, the
spontaneous  breaking of the central charge symmetry in \p{11Z} entails
half-breaking  of $N=(1,1)$ supersymmetry. This  can be also understood
from the following heuristic reasoning: once  $Z$ is spontaneously broken,
it should not give zero while applied to vacuum; then from the crossing
relation in \p{11Z} follows that $\hat{Q}_+$ or/and $\hat{Q}_-$ must also 
possess this property, i.e. it generates a spontaneously broken symmetry. 

It is worth mentioning that the transformation properties of $\Phi$ as the 
Goldstone superfield can be rederived on pure geometrical ground
proceeding from the coset (nonlinear) realizations method. Let us
identify the $N=(1,1)$ superspace co-ordinates and $\Phi$ with the parameters 
of a particular representative of the supergroup corresponding to the 
algebra \p{11Z},          
\be
G(x, \theta^+, \theta^-) = e^{i(\xp P_\pp + \xm P_=)}e^{-\theta^+\hat{Q}_+}
e^{-\theta^-\hat{Q}_-}e^{-2i\Phi(x,\theta) Z}~. \lb{coset1} 
\ee
Then the left multiplications of this element by $e^{\epsilon^+\hat{Q}_+}$ 
and $e^{\epsilon^-\hat{Q}_-}$ produce for $\Phi$ just the
$N=(1,1)$ transformations with the generators \p{QQgen}. The covariant
derivatives \p{DDgen}, \p{hatD} can be recovered from the standard Cartan
approach applied to \p{coset1}.  

Closely related to this discussion is the following phenomenon (featured by 
the $N=(2,2)$ case as well). As 
was already mentioned, the generators \p{QQgen} are defined up to a freedom 
of changing the $Z$-frame by the rotation of the type \p{simil}. In our case 
this freedom is expressed as  
\bea \lb{freedalp}
&&\hat Q, \hat D \;\Rightarrow \; \hat Q(\alpha), \hat D(\alpha) = 
e^{-\alpha \theta^+\theta^- Z}(\hat Q, \hat D)
e^{\alpha \theta^+\theta^- Z}~, \nn \\
&& \Phi \;\Rightarrow \; \Phi(\alpha) = 
e^{-\alpha \theta^+\theta^- Z}\Phi = 
\Phi -i{\alpha\over 2}\theta^+\theta^-~.
\eea
In particular, 
\be
\hat Q_+(\alpha) = Q_+ + \alpha \theta^- Z~, \quad 
\hat Q_-(\alpha) = Q_- +(2-\alpha) \theta^+ Z~. \lb{QQalpha}
\ee
We see that at $\alpha = 2$ the central charge term is entirely pumped 
over from $\hat Q_-$ to $\hat Q_+$. As a result, the 
superfield $\Phi(\alpha = 2)$ 
undergoes an inhomogeneous shift under the $N=(1,0)$ supersymmetry, i.e. 
we are facing the breaking option $N=(1,1) \rightarrow N=(0,1)$ on such 
a superfield. It can be equivalently described in terms of the $N=(0,1)$ 
components of $\Phi(\alpha = 2)$. This consideration shows that the notion 
of the linear off-shell realization of $N=(1,1)$ 
supersymmetry breaking patterns is to some extent conditional: various 
patterns are related to each other by $Z$-frame rotations 
which redefine the transformation law of the basic superfield $\Phi$ 
(any such a redefinition amounts to a constant shift of the auxiliary 
field in $\Phi$). For an arbitrary parameter $\alpha$ in \p{QQalpha} 
{\it both} fermionic fields in $\Phi$ acquire inhomogeneous pieces in their 
supersymmetry transformations, so this case corresponds to the totally 
broken $N=(1,1)$ supersymmetry off shell. This off-shell equivalence of 
various breaking options is lifted when passing on shell, or  
when going to their nonlinear realizations as explained below. 
 
Further discussion will be concentrated on the case 
$\alpha = 0$ corresponding to the pattern $N=(1,1) \rightarrow N=(1,0)$. 
The model we are considering is described by the free
action $S_1$ and as such contains no dynamics. Adding the invariant $S_2$
merely changes the algebraic equation for the auxiliary field 
$B\equiv D_+\eta^+\vert$ which is then forced to be a constant 
on shell.\footnote{Note that in the presence of such a term 
the fermionic (first) component of the superfield $\eta^+$ acquires on
shell an inhomogeneous shift under $N=(1,0)$ supersymmetry proportional to 
the constant value of auxiliary field. So in this 
case $N=(1,1)$ supersymmetry can get totally broken on shell.}
A non-trivial self-interacting model can be nonetheless constructed by
re-expressing the $N=(1,1)$ Goldstone superfield $\Phi$ in terms of the 
$N=(1,0)$ Goldstone superfield $u(x,\theta^+)$ which accommodates in a
minimal way both Goldstone degrees of freedom related to the 
spontaneously broken $\hat{Q}_-$ and $Z$ generators. This procedure  
is in a sense similar to passing from the linear sigma model with some
internal symmetry to the corresponding nonlinear sigma model.  

As the starting point, let us note that a minimal model-independent 
way to implement the $N=(1,1)$ supersymmetry spontaneously broken down 
to $N=(1,0)$ is to use the universal nonlinear realization approach 
and to introduce the Goldstone fermion superfield $\psi^- (\xp,\xm,\tp )$ 
as the coset parameter associated with the generator $\hat{Q}_-$, in full
analogy with the renowned Volkov-Akulov construction for the case of total
spontaneous breaking of supersymmetry \cite{VA}. For such a Goldstone
fermion superfield one immediately derives the universal nonlinear 
transformation law
\be\label{nlr1}
\delta\psi^-=\epsilon^- - i\epsilon^-\psi^-\partial_=\psi^- \; .
\ee
The Goldstone fermion $N=(1,0)$ superfield in any specific model of 
the spontaneous breaking $N=(1,1) \rightarrow N=(1,0)$ is expected 
to be related to $\psi^- (\xp,\xm,\tp )$ by a field redefinition. 

For the case of total spontaneous breaking of supersymmetry this
universality of the nonlinear-realization Goldstone fermion was proven in
\cite{IKa}. Also, a generic method of constructing linear representations of 
supersymmetry as nonlinear functions of the single Goldstone fermion and
its $x$-derivatives was worked out there. This approach can be
generalized rather straightforwardly to the case of partial breaking, and in
\cite{DIK} this already was done for a few simple PBGS patterns. Here we 
apply a similar construction and covariantly express the linear 
superfield representation $u, \eta^+$ in terms of the single $N=(1,0)$ 
superfield Goldstone fermion and further in terms of the basic scalar 
Goldstone superfield $u$. 

Following \cite{IKa}, \cite{DIK}, this procedure goes through two steps. 

First, we should find the {\it finite} transformation of 
the superfields \footnote{On this stage it is preferable to
deal with the superfields on which the central charge $Z$ is vanishing and
which hence contain no explicit $\theta$'s in their
supersymmetry transformations.} $\xi^- = iD_+u$ and $\eta^+$  under the
spontaneously broken $N=(0,1)$ supersymmetry with parameter $\epsilon^-$.
In our one-parameter 
case this is straightforward as the infinitesimal transformations
\p{realiz1} coincide with the finite ones:
\be      
\xi^-{}' \equiv \xi^-(\epsilon^-) = \xi^- + \epsilon^-(1 + D_+\eta^+)~,
\quad 
\eta^+{}' \equiv  \eta^+(\epsilon^-) = \eta^+ - \epsilon^-\Pm u~. 
\lb{fullsusy1}
\ee

The second step is to define new objects $\tilde{\xi}^-,
\tilde{\eta}^+$ via the substitution $\epsilon^- \Rightarrow -\psi^-$
in \p{fullsusy1}
\be\label{fullsusy2}
\tilde\xi^-= \xi^-(-\psi^-)= \xi^- -\psi^-(1+D_+\eta^+) \; ,
\quad
\tilde\eta^+= \eta^+(-\psi^-) = \eta^+ +\psi^-\partial_=u \;.
\ee

Using the transformation law \p{nlr1} of $\psi^-$,  one can check that
the superfields $\tilde\xi^-,\tilde\eta^+$ transform homogeneously and 
independently of each other under the second supersymmetry
\be
\delta\tilde\xi^-=-i\epsilon^-\psi^-\partial_=\tilde\xi^-\; , \quad
\delta\tilde\eta^+=-i\epsilon^-\psi^-\partial_=\tilde\eta^+\; .
\ee
Thus it is a covariant constraint to put these superfields equal
to zero
\be\label{constr1}
\left\{ \begin{array}{l}
\tilde\xi^-=0\\ \tilde\eta^+=0
\end{array} \right. \Rightarrow
\left\{ \begin{array}{l}
\xi^- =\psi^-\left( 1+ D_+\eta^+\right) \\
\eta^+ =-\psi^-\partial_= u~. \end{array}\right.
\ee
The system \p{constr1} can easily be solved for $\eta^+$ and $\psi^-$, 
\bea
&&\eta^+=-\frac{i D_+ u\partial_=u}{1+\Dpo\eta^+}\quad  
\Rightarrow \quad
\eta^+(u)=-\frac{2i D_+ u\partial_=u}{1+\sqrt{1-4\partial_=u
  \Pp u}}~, \label{sol1} \\
&& \psi^-(u)=\frac{2i D_+ u}{1+\sqrt{1-4\partial_=u
  \Pp u}} \; . \label{sol2}
\eea
Eq. \p{sol2} gives the anticipated equivalence relation between the
nonlinear-realization Goldstone fermion $\psi^-$ and its linear-realization
counterpart $\xi^- = iD_+u$ ($\psi^- = \xi^- + \ldots $). The
$N=(1,0)$ superfield $\eta^+$ which completes $u$ to a scalar $N=(1,1)$ 
supermultiplet is expressed by eq. \p{sol1} through $u$ itself. Thus, 
$u$ remains as the only independent quantity of our theory.

Note that the constraints \p{constr1} can be reformulated in terms of the 
$N=(1,1)$ superfield $\Phi$, but this equivalent form is not too
enlightening. We also notice that the expression \p{sol1} for $\eta^+(u)$
could be derived by imposing proper covariant constraints directly on the
covariant  derivatives $\hat{D}_{\pm}\Phi$, without introducing the
auxiliary object $\psi^-$ at the intermediate step (this would be in 
the spirit of the method of refs. \cite{Ro,RT,GPR}).
Once again, these constraints look rather involved, and it would be
difficult to guess their form. In contrast, the above universal method
unambiguously leads to the desired answer.      

Finally, we substitute the expression for $\eta^+(u)$ in both our
actions \p{action1}, \p{action2} and find that they coincide
\be\label{finaction}
S_1 = S_2= {1\over f^2}\int d^3z^{(+)}\eta^+(u) = {i\over f^2} \int d^2x
D_+ \left(\frac{D_+ u\partial_=u}{1+\sqrt{1-4\partial_=u
  \Pp u}}\right) \; .
\ee
Thus the action \p{act} with the genuinely invariant Lagrangian density 
is vanishing for $\Phi(u) = u + i\theta^- \eta^+(u)$. This can be directly 
seen from the fact that the spinor covariant derivatives of $\Phi$, eqs. 
\p{hatD}, on the shell of the constraints \p{constr1} are
proportional 
to the same Grassmann quantity 
\be
\hat{D}_-\Phi(u) = i (\theta^- + \psi^-) \Pm u~, \quad 
\hat{D}_+\Phi(u) = -i (\theta^- + \psi^-)(1 +D_+\eta^+)~,
\ee  
and so the Lagrangian density in \p{act} equals to zero for $\Phi(u)$.

In accord with the general concept of the nonlinear-realization method we 
expect that the Goldstone superfield action \p{finaction} is universal, in
the sense that it describes the low-energy dynamics of any $D=2$ 
model where a spontaneous breaking of $N=(1,1)$ supersymmetry down to
$N=(1,0)$ with a scalar Goldstone multiplet occurs. To reveal 
its relation to string theory, let us note that   
for the physical bosonic component $u_0=u|_{\theta^+ =0}$ one obtains just
the static-gauge form of the Nambu-Goto $D=3$ string action
\be
S_{bos}=\frac{1}{4}\int d^2x 
\left(1 - \sqrt{1-4\partial_=u_0 \Pp u_0}\right)~.
\ee
This suggests an interpretation of the Goldstone field $u$ as the
transverse string co-ordinate and of the whole superfield action
\p{finaction} as the static-gauge form of the action of the $N=1, D=3$
superstring in a flat Minkowski background. Actually, from the $D=3$
perspective the superalgebra \p{11Z} is just the $N=1$ Poincar\'e
superalgebra, with $Z$ being the momentum in the third direction. 
The component form of the
action \p{finaction} could be recovered from the standard Green-Schwarz 
action for this superstring, like it has been done in \cite{HP} for the
PBGS form of the $N=1, D=4$ superstring action (we shall reproduce this
example in Sect. 5). A novel point of our consideration is that 
the action \p{finaction} was constructed directly in $D=2$ superspace,
proceeding only from the purpose to describe the partial breaking $N=(1,1)
\rightarrow N=(1,0)$ and not assuming any $D=3$ structure beforehand. 

As a final remark, note that we could equally start from 
the linear realization \p{QQalpha} with $\alpha = 2$ which corresponds 
to the partial breaking option $N=(1,1) \rightarrow N=(0,1)$. The relevant 
nonlinear realization is constructed along the same lines, but with 
$u(x, \theta^-) = \Phi(\alpha = 2)|_{\theta^+ = 0}$ as the irreducible
Goldstone superfield. As a $D=2$ field theory, it is clearly non-equivalent 
to the previous one because the Goldstone fermions in both theories have 
opposite light-cone chiralities (purely bosonic sectors are identical).  
Nevertheless, their Goldstone superfield actions are related to each other 
by a kind of mirror symmetry and are gauge-equivalent from the $D=3$ 
perspective, corresponding to two different choices of gauge with respect 
to kappa-symmetry in the same $N=1, D=3$ superstring Green-Schwarz action. 
A nonlinear realization of the total breaking pattern \p{QQalpha} 
with $\alpha \neq 0, 2$ can be straightforwardly constructed by 
the original methods of \cite{IKa}. It is a modification of the standard 
Volkov-Akulov theory for $N=(1,1), D=2$ \cite{Ro}, with one extra Goldstone 
scalar field for the spontaneously broken central charge
generator.\footnote{Such a modification was considered in \cite{victor}.}
Thus various options of the $N=(1,1)$ 
breaking,  being equivalent modulo $Z$-frame rotations at the level of the 
linear realization, yield non-equivalent $D=2$ theories after passing 
to the relevant nonlinear realizations.

\subsection{$N=(2,0)\rightarrow N=(1,0)$}

This case is somewhat special. The main peculiarity is that the $(2,0)$
superalgebra
\be\label{n20}
\left\{ Q_+^1,Q_+^1\right\}=\left\{ Q_+^2, Q_+^2 \right\} = 2P_\pp \;
\ee
admits no $SO(1,1)$- scalar central charges. Hence, we do not
expect to find a scalar bosonic field with a shift symmetry among the set
of our fields. The generators $Q_+^{1,2}$ and covariant 
derivatives $D_+^{1,2}$ are defined by the standard formulas without 
central charge terms, 
\be
Q^{1,2}_+ = \frac{\partial}{\partial \theta^+_{1,2}} +
i\theta^{+}_{1,2}\Pp~, \quad                       
D^{1,2}_+ = \frac{\partial}{\partial \theta^+_{1,2}} -
i\theta^{+}_{1,2}\Pp~.
\ee

We wish to describe the situation where $N=(1,0)$ supersymmetry generated
by $Q_+^1$ is unbroken while the other $N=(1,0)$ supersymmetry, with generator 
$Q_+^2$, is spontaneously broken. Thus, like in the preceding 
Subsection, we are led to introduce a real Goldstone fermionic $N=(1,0)$ 
superfield 
\be
\xi^+(\xp, \xm, \theta^+_1) = \lambda^+(\xp, \xm) + \theta^+_1 F(\xp,
\xm) \lb{compxi}
\ee
($[\xi^+] = -1/2$), which contains a chiral fermion $\lambda^+$ and an
auxiliary real bosonic
field $F$. As opposed to the previously discussed case we cannot represent
$\xi^+$ as a covariant spinor derivative of some scalar $N=(1,0)$
superfield since we now have at our disposal only one spinor derivative 
$D^1_+$ ($(D^1_+)^2 = -i\Pp$). Nevertheless, we can proceed in a similar
way and, first of all, try to construct a linear realization of
the breaking pattern by extending $\xi^+$ to an $N=(2,0)$
multiplet. The simplest possibility is to add one more fermionic superfield
$\eta^+(\xp,\xm,\tp )$ which can be combined with $\xi^+$ to an 
$N=(2,0)$ supermultiplet. The following transformations  
\be\label{realiz2}
\delta\xi^+=\epsilon^+_2 +\epsilon^+_2D_+^1\eta^+\:\; , \quad
\delta\eta^+=-\epsilon^+_2D_+^1\xi^+  
\ee 
can be checked to form just the $Q^2_+$ part of the algebra \p{n20}.
These $N=(1,0)$ superfields are related to a chiral spinor $N=(2,0)$ 
superfield $\Xi^+$ via  
\be
\Xi^+ (\xpl, \xm, \theta^+_1 + i\theta^+_2) = \xi^+ - i\eta^+ + 
i\theta^+_2 D^1_+(\xi^+ - i\eta^+)~, \lb{Ksi}
\ee  
where we expanded $\Xi^+$ in $\theta^+_2$ by making use
of the definition of $\xpl$ in \p{A17}. Assuming for $\Xi^+$ the
inhomogeneous
transformation law under the $\epsilon^+_2$ supersymmetry
\be
\delta \Xi^+ = \epsilon^+_2 + (\epsilon^+_2 Q^2_+)\,\Xi^+~, 
\ee
we get for $\xi, \eta$ just the transformation laws \p{realiz2}. 

Like in the previous case, one can construct for this supermultiplet 
two independent off-shell invariants
\bea
&&S_1 = {1\over 2f^2} \int d^4z^{(+)}\Xi^+\bar\Xi^+ = 
-{i\over f^2}\int d^3z^{(+)}\left(\xi^+D^1_+\xi^+ + 
\eta^+D^1_+\eta^+ \right)~, \lb{invN21} \\
&&S_2 = -{1\over 2\sqrt{2} f^2}(b +i) \int d\xm d^2\zeta^{(+)}_L \Xi^+ +
\mbox{c.c.} = {i\over f^2} \int d^3z^{(+)} (\eta^+ + b \xi^+)~, \lb{invN22}   
\eea 
where $b$ is an arbitrary real constant. The first invariant describes the
free theory of two chiral fermions. Adding the second
invariant merely changes the algebraic equations of motion for the
auxiliary fields $D^1_+\eta^+\vert$ and $D^1_+\xi^+\vert$ allowing them to 
be non-zero constants. In its presence, the on-shell pattern of
supersymmetry breaking differs from the off-shell one we started with. 
In particular, the entire $N=(2,0)$ supersymmetry can be broken.  

Once again, it is possible to trade the second superfield $\eta^+$ for the 
Goldstone one $\xi^+$ to obtain the minimal theory in terms of the 
single superfield $\xi^+$ with a nonlinearly realized $\epsilon^+_2$
supersymmetry. To this end, we construct the superfields 
\be
\tilde{\xi}^+ =\xi^+ -\psi^+(1 + D^1_+\eta^+)~, \quad \tilde{\eta}^+
=\eta^+ + \psi^+ D^1_+\xi^+~, 
\ee 
where $\psi^+(x, \theta_1^+)$ is the Goldstone fermion with the universal
transformation law 
\be
\delta\psi^+=\epsilon^+_2 - i\epsilon^+_2\psi^+\Pp\, \psi^+
\ee
(cf. eq. \p{nlr1}). These objects transform homogeneously,
$$
\delta \tilde{\xi}^+ = - i\epsilon^+_2\psi^+\Pp\, \tilde{\xi}^+~, \quad 
\delta \tilde{\eta}^+ = - i\epsilon^+_2\psi^+\Pp\, \tilde{\eta}^+~,
$$ 
and hence can be covariantly equated to zero, leading to the desired 
expressions of both $\psi^+$ and $\eta^+$ in terms of $\xi^+$:  
\bea
&& \tilde{\xi}^+ = \tilde{\eta}^+ = 0 \quad \Rightarrow \nn \\
&& \eta^+ = - \frac{2\,\xi^+D^1_+\xi^+}{1 +\sqrt{1 -4(D^1_+\xi^+)^2}}~,
\quad 
\psi^+ = \frac{2\,\xi^+}{1 + \sqrt{1 -4(D^1_+\xi^+)^2}}~. \lb{expr21}
\eea

Substituting this expression for $\eta^+$ into the invariants \p{invN21}
and \p{invN22}, we find that for $b=0$ they coincide: 
\be\label{action3a} 
S_1=S_2(b=0)=-{2i\over f^2} \int d^3z^{(+)}
\left(\frac{\xi^+D^1_+\xi^+}{1+\sqrt{1-4\left( D^1_+\xi^+
\right)^2}}\right)
\;. 
\ee 
Adding the term $\sim b\neq 0$ results only in a modification 
of the equation of motion for the non-propagating field 
$F = D^1_+\xi^+\vert$ which becomes a constant $\sim b$ 
on shell. As a result, 
the Goldstone component $\lambda^+$ starts to  transform inhomogeneously
under the $(1,0)$ supersymmetry which was originally unbroken off shell. 
Clearly, there still exists a combination of the supersymmetry generators
under which $\lambda^+$ transforms homogeneously, so the effect of partial
breaking retains on shell. However, only in the $b=0$ case the off- and
on-shell patterns of this breaking are in one-to-one correspondence.   

Despite its nonlinear appearance, the Goldstone superfield action
\p{action3a} in the present case yields trivial dynamics. The component 
form of \p{action3a} is as follows  
\be 
S_1 \sim  
\frac{i\lambda^+ \Pp\, \lambda^+}{(1 + \sqrt{1-4F^2})\sqrt{1-4F^2}} + 
{1\over 4}(1 - \sqrt{1-4F^2})
\lb{comp3a}
\ee
(the component fields were defined in \p{compxi}). It gives rise to the free 
equations for the involved fields 
\be\label{eom2a}
\partial_\pp\, \lambda^+=0\; , \quad F=0 \:.
\ee
The absence of interaction can be made manifest by making the
invertible field redefinition 
\be
\mu^+ \equiv \frac{\sqrt{2}\;\xi^+}{\sqrt{1+\sqrt{1-4\left( D^1_+\xi^+
\right)^2}}}~, \quad \xi^+ = \mu^+\,\sqrt{1 - \left(D^1_+\mu^+ \right)^2}~,
\ee
after which the action \p{action3a} is reduced to the free one~,  
\be\label{actionfree} 
S_1 =-{i\over f^2}\int d^3z^{(+)} \mu^+D^1_+\mu^+
\;. 
\ee

Thus, the $N=(2,0)\rightarrow N=(1,0)$ PBGS in $D=2$ corresponds to a
degenerate system of one free left-chiral fermion $\lambda^+(x)$
($\lambda^+ = \lambda^+(\xm)$ on shell).

It is worth mentioning that, in accord with the well-known $D=2$ 
equivalence between the abelian gauge field strength and an auxiliary
field, the fermionic superfield $\xi^+$ can also be treated as a covariant 
field strength for the $N=(1,0)$ fermionic and bosonic ``gauge potentials'' 
$A_+$ and $A_=$ \cite{Gates}:
\be
\xi^+ \;\Rightarrow \; \xi^+_v = \Pm A_+ - D^1_+ A_=~, \;\; 
\delta A_+ = D^1_+ \Lambda~, \; \delta A_= = \Pm \Lambda~.         
\ee
In the Wess-Zumino gauge $\xi_v^+$ has the same component content as in  
\p{compxi}, with the auxiliary $F$ substituted by the field strength 
of the $D=2$ vector gauge field $(v_\pp\;, v_=)$
\be
F \;\Rightarrow \; F_v = \Pp\, v_= - \Pm v_\pp~. \lb{FFv}
\ee
The supersymmetry transformation properties of $\xi^+_v$ do not change 
compared to those of $\xi^+$, and the invariant action is still
given by \p{action3a} and \p{comp3a}. The action of the fermionic field 
becomes free after a proper field rescaling, while the $F_v$ part is
recognized as the $D=2$ Born-Infeld action 
\be
S(F_v) \sim \int d^2x\, \left(1 - \sqrt{1-4F_v^2}\right)~. \lb{BI20}    
\ee
Thus the model of partial breaking $N=(2,0)\rightarrow N=(1,0)$ 
with the $N=(1,0)$ vector Goldstone multiplet amounts to a sort of 
``space-filling'' $N=(2,0)$ D1-brane, which has just this multiplet as
the physical world-sheet one. Since the gauge field is non-dynamical in
$D=2$ at the classical level irrespective of its action, we are still left
with a free theory of one chiral fermion in this case. 

Note that at the quantum level or for non-vanishing first Chern number the
abelian gauge field is non-trivial \cite{Wit1}. In
this connection, it is worthwhile to mention that the term $\sim b$ in
\p{invN22} yields just the well-known topological invariant of the $D=2$
abelian gauge field upon the substitution  \p{FFv}. The chiral $N=(2,0)$
superfield $\Xi^+$ \p{Ksi} with $\eta^+$ covariantly expressed through 
$\xi_v^+$ by \p{expr21} can be identified with the superfield  strength of 
the $N=(2,0)$ gauge vector multiplet  \cite{Gates,HPS,DS}
\be
\Xi^+_v (x, \theta^+) = \Pm\bar D_+ V - \bar D_+ A_=~, \lb{20str} 
\ee
where the real $N=(2,0)$ superfield $V$ and antichiral superfield $A_=$, 
$D_+A_= = 0$, are the corresponding gauge potentials 
$$
\delta V = a + \bar a~, \quad \delta A_= = \Pm a~, \quad D_+ a =0~.
$$
The coefficient in front of the $N=(2,0)$ superspace integral in
\p{invN22} is recognized as the familiar complex coupling constant
\cite{Wit1,MP} combining the coefficient of the Fayet-Iliopoulos 
term ($\eta^+_v$ in \p{invN22}) and the well-known $\theta$-angle: 
$$
{1\over f^2} (i + b) \equiv \tau = ir + {\theta\over 2\pi}~.
$$  
Thus the parameter $b$ has the physical meaning of a $\theta $-angle.

Finally, let us notice that all models of this and the following Sections 
submit to the argument used in \cite{HP,HLP} to evade the partial breaking
no-go theorem of \cite{Wit2}. Namely, the Noether energy-momentum tensors
corresponding to the $D=2$ translation generators in the algebras of broken
and unbroken supersymmetries do not coincide, but differ by some constants. 
These constant ``central charges'' should not be confused with the active 
central charges which act as shifts of the Goldstone superfields and
the presence of which in many cases turns out to be crucial for triggering 
the partial spontaneous breakdown of supersymmetry. Also, 
to avoid a possible misunderstanding, we point out that all superalgebras
used throughout this paper should be regarded as algebras of infinitesimal 
field transformations rather than as algebras of the charge and supercharge
generators computed by the Noether procedure from some invariant
action. For spontaneously broken symmetries and supersymmetries 
the latter  objects are often ill-defined, while the algebras of the
corresponding currents and supercurrents are always meaningful.      
  
\setcounter{equation}0
\section{\lb{B} Partial breaking N=(2,2) to N=(1,1)}

We start the study of the partial breaking pattern $N=(2,2)\rightarrow
N=(1,1)$ by constructing its linear realization.  Without loss of generality, 
we let $Q^1_{\pm}$ generate the unbroken $N=(1,1)$ supersymmetry. 

A simple analysis shows that the scalar real
$N=(1,1)$ superfield $\pi(x, \theta^+_1, \theta^-_1)$ has the components
content just appropriate for the relevant Goldstone supermultiplet,
including two real fermionic fields $\kappa^\pm(x)$ capable to 
be the Goldstone fermions associated with the spontaneously broken 
$Q^2_{\pm}$ supersymmetry: 
\be
\pi=p(x)+i\tpo\kappa^-(x)+i\tmo\kappa^+(x)+i\tpo\tmo F(x)~.\lb{B7}
\ee
Let us show that this multiplet can be naturally embedded into 
a chiral scalar $N=(2,2)$ superfield $\phi(\xpl, \xml, \theta^+,
\theta^-)$ with the appropriate transformation law including
central-charge terms. It is similar to the 
transformation laws \p{PhiZ}, \p{transPhi} of the Goldstone superfield 
\p{1decom} in the linear realization of the partial breaking
$N=(1,1)\rightarrow N=(1,0)$ considered in Subsect. 3.1.  

We shall need the structure of the $N=(2,2)$
supersymmetry generators with central charges in the realization on 
chiral superfields. Recalling the discussion after eq. \p{A21}, we are led  
to choose the frame where the spinor derivatives $\bar D_\pm$ contain no
central charge terms and where the standard chirality constraints 
\p{A20} are valid. Bearing also in mind that in $N=(2,2)$ supersymmetry 
with central charges the chiral superfields can be defined only under 
the restriction $Z_3 -iZ_4 = 0$ in \p{A8} - \p{explicD} and \p{A9b} -
\p{q1alg2} (recall eq. \p{chirint}), the explicit form of the $N=(2,2)$
generators in this frame 
is as follows 
\bea 
\hat{Q}_+ &=& \frac{\partial}{\partial \theta^+} 
+i \bar\theta^+ \Pp - \theta^- (Z_3 +iZ_4)  \equiv 
Q_+ - \theta^- (Z_3 +iZ_4)~, 
\nn \\ 
\hat{\bar Q}_+ &=& \frac{\partial}{\partial \bar\theta^+} +i \theta^+ \Pp 
+ 2\theta^- (Z_1 -iZ_2) \equiv \bar Q_+ + 2\theta^- (Z_1 -iZ_2)~, \nn \\
\hat{Q}_- &=& \frac{\partial}{\partial \theta^-} +i \bar\theta^- \Pm 
- \theta^+(Z_3 +i Z_4) \equiv Q_- - \theta^+(Z_3 +i Z_4)~,\nn \\
\hat{\bar Q}_- &=& \frac{\partial}{\partial \bar\theta^-} +i
\theta^- \Pm  +2\theta^+ (Z_1 +iZ_2) \equiv \bar Q_- + 2\theta^+ (Z_1 +iZ_2)~,
\lb{explicQ} \\
\delta \phi &=& (\epsilon \hat{Q})\phi = (\epsilon Q)\phi + 
2\bar\epsilon^{\,+}\theta^-(Z_1-iZ_2)\phi + 2\bar\epsilon^{\,-}\theta^+
(Z_1+iZ_2)\phi  \nn \\
&& - \, (\epsilon^+\theta^- + \epsilon^-\theta^+)(Z_3 +iZ_4)\phi~. 
\lb{tranphi}
\eea     

Like in the $N=(1,1) \rightarrow N=(1,0)$ case, in order to trigger 
the spontaneous breaking of $N=(2,2)$ to $N=(1,1)$ the superfield $\phi$ is
expected to undergo pure shifts under the action of the central charge 
generators $Z_1, Z_2$. Since $\phi$ is complex, it can be shifted by a
complex parameter, i.e. in principle both $Z_1$ and $Z_2$ can be
non-vanishing on $\phi$. However, we wish to have the $N=(1,1)$ generators  
$$
\hat{Q}^1_\pm \sim \hat{Q}_\pm + \hat{\bar Q}_\pm = Q_\pm + \bar Q_\pm 
-\theta^\mp (Z_3 + iZ_4) + 2 \theta^\mp (Z_1 \mp iZ_2) 
$$
unbroken, i.e. having 
no central charge terms. We still have a freedom of changing the $Z$-frame 
according to the rule \p{simil},
\be 
(D, \hat{Q}) \;\; \Rightarrow \;\; e^{-A}(D, \hat{Q})e^A\;, \quad 
\phi \;\; \Rightarrow \;\; \hat\phi = e^{-A}\phi \lb{freed}
\ee
with $A \sim \theta^+\theta^- \times [\mbox{central charges}]$. Such a
transformation commutes with $\bar
D_\pm$ and so does not affect the chirality condition \p{A20} (it 
amounts to a redefinition of the auxiliary field in $\phi$). At the same
time it gives rise to the appearance of the central charge terms in the 
generators $\hat{Q}_\pm $, leaving intact $\hat{\bar Q}_\pm$. A simple 
inspection shows that the necessary and sufficient conditions for 
the central charge terms to drop out from $\hat{Q}^1_\pm $, modulo the
freedom \p{freed}, is the following one, 
\be
[2Z_1 - (Z_3 +i Z_4)]\phi = 0 \lb{Z1cond}       
\ee
(together with \p{chirint}, it implies $Z_1 = Z_3$).
Then, the rotation \p{freed} with 
\be
A = 2 i\,\theta^+\theta^-\, Z_2 \lb{A2211}
\ee
eliminates the central charge terms from $\hat{Q}^1_\pm $. 

There exist several possibilities to realize the breaking pattern 
$N=(2,2) \;\rightarrow \; N=(1,1)$ on $\hat\phi$. In particular, one 
could keep two independent central charges producing a complex shift
of $\hat\phi$, with two Goldstone-type $N=(1,1)$ supermultiplets. We are
interested in the minimal possibility with a single Goldstone 
multiplet comprised by the $N=(1,1)$ superfield \p{B7}. The choice of 
central charges giving rise to this option can be shown to be as follows,  
\bea
&&Z_1\,\phi = Z_3\,\phi = Z_4\,\phi = 0~, \quad  
Z_2\,\phi = -{1\over 2\sqrt{2}}~, \lb{Z1cond2} \\       
&&\delta_z \phi = 
-2\sqrt{2} ia \left[Z_2, \;\phi\right] = ia~. 
\lb{Z2cond}  
\eea
Then, for the rotated chiral superfield  
\be
\hat{\phi} = e^{-A}\phi = \phi - 2i\theta^+\theta^- Z_2\,\phi = \phi + 
{i\over \sqrt{2}} \theta^+\theta^- 
\ee
we get the following transformation law
\be
\delta \hat{\phi} = (\epsilon \hat{Q})\hat{\phi} = \epsilon^+_2\theta^- -
\epsilon^-_2\theta^+ +  
(\epsilon Q)\hat{\phi}~. \lb{fintran} 
\ee
Here the generators $\hat{Q}$ are related to the original ones 
\p{explicQ} by the rotation \p{freed} with $A$ \p{A2211} (using the 
same notation for these two sets of generators will hopefully 
not lead to a confusion). 

A few remarks are in order at this stage.

Firstly, let us point out that in the presence of all 
central charge generators in \p{explicQ} acting as shifts 
of $\phi$, only one real combination of the $N=(2,2)$ supercharges 
can be arranged to include no central charge terms by performing an 
appropriate $Z$-frame rotation. As a result, in this
case there becomes possible the 1/4 partial breaking of $N=(2,2)$ 
supersymmetry down to its $N=(1,0)$ (or $N=(0,1)$) subgroup, 
with three out of four real fermionic component fields of $\phi$ 
having inhomogeneous transformation laws and so being Goldstone fermions. 
This option will be considered in Section 6.

Secondly, we could choose 
$$
Z_2\,\phi = 0~, \quad Z_1\,\phi \neq  0~, \quad (Z_3+iZ_4)\phi = 0    
$$
instead of \p{Z1cond2}. In this case it is possible to remove 
central charge terms from the $N=(1,1)$ generators $\hat{Q}^1_+~, \,
\hat{Q}^2_- \sim i(\hat{Q}_- - \hat{\bar Q}_-)$ (or $\hat{Q}^1_-~, \,
\hat{Q}^2_+$). This option amounts to an equivalent pattern of the partial
breaking of $N=(2,2)$ supersymmetry to $N=(1,1)$. 

As a last remark we note that 
it is possible to construct the real $N=(2,2)$ superfield basically as a
sum of $\hat{\phi}$ and $\hat{\bar\phi}$. It is invariant under $Z_2$
shifts and, as a consequence, has the standard homogeneous transformation
properties under the full $N=(2,2)$ supersymmetry. To construct it, one
should make the similarity rotation back to the central-basis frame of
$N=(2,2)$ superspace where the generators $\hat{Q}_\pm$ and $\hat{\bar
Q}_\pm$ become conjugated to each other in the ordinary sense. The chiral
$N=(2,2)$ superfield is ``covariantly chiral'' in such a frame. The 
precise relation between two frames is given by  
\be
\phi_{cov} = e^B \hat{\phi} = \hat{\phi} +{i\over
2\sqrt{2}}(
\theta^+\bar\theta^- + \bar\theta^+\theta^- - 2\theta^+\theta^-)~, 
\ee
with 
$$
B = -i(\theta^+\bar\theta^- + \bar\theta^+\theta^- - 2\theta^+\theta^-)\,Z_2~.
$$   
It is straightforward to check that the real superfield  
\be
\hat{\Phi} \equiv  2 Re\,\phi_{cov} = \hat{\phi} + \hat{\bar\phi} 
+i \sqrt{2}\, \theta^+_2\theta^-_2  \lb{B0b} 
\ee  
possesses zero central charge and transforms in the conventional way under
$N=(2,2)$ supersymmetry
\be   
\delta \hat{\Phi} = (\epsilon Q)\,\hat{\Phi}~.
\ee
Here $Q_\pm, \bar Q_\pm$ contain no central charge terms.

In our further exposition we shall closely follow the lines we 
pursued for the toy examples in the previous Section.  

We need to know how $\hat{\phi}$ is expressed through 
$N=(1,1)$ superfields appearing in its $(\tpt,\tmt)$ expansion. 
The chirality conditions \p{A20} rewritten in the real basis
\be
(\Dpo+i\Dpt )\phi=(\Dmo+i\Dmt )\phi=0\lb{B1}
\ee
can be solved in terms of the complex $N=(1,1)$ superfield $\chi=
{1\over \sqrt{2}}(\sigma+i\pi)$ as follows 
\be
\phi(x_L,\theta^\pm)= {1\over \sqrt{2}}\left[1+i\tpt\Dpo
+i\tmt\Dmo+ \tpt\tmt\Dpo\Dmo \right]\left[\sigma (x,\theta_1^\pm) 
+ i\pi (x,\theta^\pm_1) \right]
~.\lb{B2}
\ee

The infinitesimal $N=(2,2)$ superfield transformation \p{fintran} implies,
via the relation \p{B2}, the following transformation laws for the real
$N=(1,1)$ superfields $\sigma$, $\pi$:
\bea
&&\delta \pi=i(\emt\tpo -\ept\tmo )+(\ept\Dpo+\emt\Dmo)\sigma~,\nn\\
&&\delta \sigma=-(\ept\Dpo+\emt\Dmo)\pi~.\lb{B3}
\eea
In the closure of these transformations one finds a pure shift of $\pi$
generated by $Z_2$, so $\pi$ can be interpreted as the $N=(1,1)$
Goldstone superfield for the spontaneously broken central charge
transformations. Its spinor derivatives, 
\be
\xi^- \equiv  i D^1_+\pi~, \qquad \xi^+ \equiv -i D^1_-\pi~, \lb{gf21}
\ee
also transform inhomogeneously 
\be
\delta \xi^- = \epsilon^-_2 (1 -iD^1_+D^1_-\sigma) -\epsilon^+_2\Pp\,
\sigma~, \;
\delta \xi^+ = \epsilon^+_2 (1 -iD^1_+D^1_-\sigma) +\epsilon^-_2\Pm
\sigma~,  \lb{B33}   
\ee
and so they are the relevant linear-realization Goldstone fermions for 
the spontaneously broken half of $N=(2,2)$ supersymmetry.

Like in the previous examples, in this linear realization 
all invariants one can construct correspond to a free theory. In
the $N=(2,2)$ and $N=(1,1)$ superspaces they are represented by the
following expressions   
\bea
&&S_1= {1\over 2f^2}\int d^6z\,\phi\bar\phi= -{i\over f^2}\int d^4z
(\Dpo\sigma\Dmo\sigma+\Dpo\pi\Dmo\pi)~,\lb{B3e} \\
&& S_2 = -\frac{(b + i)}{\sqrt{2} f^2}\int d^4\zeta_L\;\phi
+\mbox{h.c.}= {1\over f^2}\int d^4z (\sigma +b\pi)~.\lb{B3d}
\eea

The next step is to construct the nonlinear realization of the spontaneous 
breaking $N=(2,2) \rightarrow N=(1,1)$ in terms of the sole  
scalar Goldstone multiplet $\pi (x, \theta_1^\pm)$. As before, 
we should firstly construct the $N=(1,1)$ superfields with a homogeneous
transformation law under the spontaneously broken supersymmetry.
This can be done according to the universal prescription, by 
substituting the universal nonlinear realization Goldstone fermion 
$N=(1,1)$ superfields $\psi^\pm$ for the supergroup parameters in
the finite transformation laws. These Goldstone fermions transform 
according to  
\bea
\delta \psi^+ = \epsilon^+_2 -i (\epsilon^+_2\psi^+\Pp + 
\epsilon^-_2\psi^-\Pm)\psi^+~, \quad 
\delta \psi^- = \epsilon^-_2 -i (\epsilon^+_2\psi^+\Pp + 
\epsilon^-_2\psi^-\Pm)\psi^-~.
\eea

Denoting the finite transforms of $\xi^\pm$ and $\sigma$ by 
$\tilde{\xi}^\pm(\epsilon^+_2, \epsilon^-_2)$,
$\tilde{\sigma}(\epsilon^+_2,
\epsilon^-_2)$ (these quantities can be straightforwardly
constructed by the infinitesimal laws \p{B3}, \p{B33}), the constraints
read 
\be
\tilde{\xi}^\pm (-\psi^+, -\psi^-) = 0~,\qquad 
\tilde{\sigma} (-\psi^+, -\psi^-) = 0~. \lb{2211constr}
\ee 
Once again, these are covariant with respect to both unbroken
and broken supersymmetries since the $N=(1,1)$ superfields on their l.h.s. 
transform homogeneously under the $\epsilon_2$ transformations:
$$
\delta \tilde{\xi}^\pm = -i (\epsilon^+_2\psi^+\Pp + 
\epsilon^-_2\psi^-\Pm)\tilde{\xi}^\pm~, \quad 
\tilde{\sigma} = -i (\epsilon^+_2\psi^+\Pp + 
\epsilon^-_2\psi^-\Pm)\tilde{\sigma}~.
$$ 

Explicitly, eqs. \p{2211constr} are as follows 
\bea
&&\Dpo \pi+i\psi^- \left(1 - i\Dpo\Dmo\sigma\right) -i\psi^+\Pp\sigma
-i\psi^+\psi^-\Pp\Dmo\pi=0~,\nn\\
&&\Dmo \pi-i\psi^+\left(1 -i\Dpo\Dmo\sigma\right)-i\psi^-\Pm\sigma
+i\psi^+\psi^-\Pm\Dpo\pi=0~,\lb{lino} \\
&& \sigma+\psi^+\Dpo \pi+\psi^-\Dmo
\pi+i\psi^+\psi^-(1-i\Dpo\Dmo\sigma)=0~.\lb{sigcom}
\eea 
The first two equations set the equivalence relation between the linear- 
and nonlinear-realizations Goldstone fermions, while the third one (most
essential) serves to nonlinearly express the superfield $\sigma$ through 
the only independent remaining superfield, the scalar Goldstone 
superfield $\pi$. Note that the linear-realization 
Goldstone fermions $\sim D^1_\pm\pi$ obey the obvious integrability
condition $D^1_+(D^1_- \pi) = -D^1_-(D^1_+\pi)$, so their 
counterparts $\psi^\pm$, in virtue of the relations \p{lino}, also 
obey some covariant condition reducing their component content to 
that of a scalar $N=(1,1)$ multiplet. We should not care about this 
since $\psi^\pm$ are auxiliary objects which appear only at the 
intermediate step. 

The expression for $\sigma(v)$ can be found rather easily, 
observing that the relations
\p{lino}
imply 
\bea
&& i\psi^+\psi^-(1-i\Dpo\Dmo\sigma) = - \psi^+\Dpo \pi = -\psi^-\Dmo \pi
\quad \Rightarrow \quad \nn \\
&& \sigma(v) = - \psi^+\Dpo \pi = -\psi^-\Dmo \pi = 
-{1\over 2}\left(\psi^+\Dpo \pi + \psi^-\Dmo \pi \right)~. \lb{B5c}  
\eea
Then simple manipulations using the nilpotency of the fermionic
superfields give
\be
\sigma(\pi) = -\frac{i\,D^1_+\pi D^1_-\pi}{1 - iD^1_+D^1_-\sigma}~.
\ee
From this relation it is easy to find the ``effective'' part of 
$D^1_+D^1_-\sigma $ containing no nilpotent quantities $D^1_\pm\pi$ 
and to obtain the final expression    
\be
\sigma(\pi)= -\frac{2i\,\Dpo\pi\Dmo\pi}{1+\sqrt{1-4 W}}~,\q
W=\Dpo\Dmo
(\Dpo\pi\Dmo\pi)~.\lb{B5}
\ee

Now we can treat the transformation
\be
\delta \pi=i(\emt\tpo -\ept\tmo )+(\ept\Dpo+\emt\Dmo)\sigma(\pi)\lb{vreal}
\ee
as a specific form of the nonlinear realization and construct the
Goldstone $N=(2,2)$ superfield $\hat\phi$ in this parametrization,
\be
\hat\phi(\pi)={1\over \sqrt{2}}\left[1+i\tpt\Dpo
+i\tmt\Dmo+\tpt\tmt\Dpo\Dmo\right]\left(\sigma(\pi) + i\pi \right)~.
\ee

It is worth noting that the same nonlinear realization of 
the partial breaking $N=(2,2) \rightarrow N=(1,1)$ can be recovered 
by imposing, in the spirit of refs. \cite{RT,GPR}, the nilpotency condition 
on the homogeneously transforming real $N=(2,2)$ superfield $\hat\Phi$
defined in \p{B0b}, 
\be 
\hat\Phi^2 = 0~. \lb{nilp}
\ee
In terms of $N=(1,1)$ superfields $\pi$ and $\sigma$ this $N=(2,2)$ 
superfield  is expressed as 
\be
\hat{\Phi} \sim \sigma - \theta^+_2 D^1_+\pi - \theta^-_2 D^1_-\pi 
+ i\theta^+_2\theta^-_2 (1 -i D^1_+D^1_-\sigma)
\ee
and constraint \p{nilp} implies for them that
$$
\sigma^2 = 0~, \quad \sigma D^1_\pm \pi = 0~, \quad 
\sigma = - \frac{i\,D^1_+\pi D^1_-\pi}{1 - iD^1_+D^1_-\sigma}~.  
$$

Let us turn to constructing invariant actions for the Goldstone 
superfield $\phi$. It is straightforward to check that 
\be
D^1_+\sigma(\pi)D^1_-\sigma(\pi) + D^1_+\pi D^1_+\pi =i\sigma(\pi)~. 
\ee  
Hence, similarly to the previous cases, the invariants $S_1$ and
$S_2$ in \p{B3e} and \p{B3d} basically coincide with each other, 
\be
S_1 = S_2 (b=0) = {1\over f^2}\int d^4z \sigma
(\pi) = -{2i \over f^2} \int d^4 z 
\left(\frac{\Dpo\pi\Dmo\pi}{1+\sqrt{1-4 W}} \right)~.\lb{B6}
\ee 
The most general Goldstone superfield action includes a non-zero parameter
$b$,
\be
S = S (b) = {1\over f^2}\int d^4z \left(\sigma(\pi) +
b\,\pi\right)~.\lb{B66}
\ee 
As opposed to the toy example of breaking $N=(2,0) \rightarrow N=(1,0)$ 
(Sect. 3.2), this action necessarily contain a nonpolynomial
self-interaction, so adding the term $\sim b$ can have an impact  
on the associated dynamics. However, in the presence of such a term the
original unbroken $N=(1,1)$ supersymmetry gets broken on shell: the 
auxiliary field $F = iD^1_+D^1_-\pi \vert$ acquires a non-zero 
constant part $\sim b$, thus giving rise to inhomogeneous pieces in
the $\epsilon_1$ transformations of the Goldstone fermions $\kappa^\pm$. 
Though on shell there still exist two unbroken linear combinations of the
$N=(2,2)$ generators, the off- and on-shell patterns of the partial
breaking prove to be different. If we wish these options to coincide,
the term $\sim b$ can be ignored.

To establish links with string theory, let us examine the bosonic
sector of the Goldstone superfield action \p{B6}:
\be
S_1^{bos} = {1\over 2f^2} \int d^2x \left[1 -
\sqrt{1 -4(\Pp\,p\Pm p + F^2)}\right]~. \lb{boson21}
\ee       
The equation of motion for the auxiliary field $F$ implies 
\be
F = 0~.
\ee 
After substituting this into \p{boson21}, the latter is reduced to the
static-gauge Nambu-Goto action for a string in $D=3$ Minkowski space, 
with $p(x)$ being the corresponding transverse coordinate. From the $D=3$
standpoint, the $N=(2,2)$, $D=2$ superalgebra  \p{A23c} with $Z_1 = Z_3 =
Z_4 = 0$ is the $N=2$ Poincar\'e superalgebra, with $Z_2$ completing
the pair $(P_=,\; P_\pp)$ to the full $D=3$ energy-momentum vector.
Indeed, combining the real generators $Q^1_\pm, Q^2_\pm$ into two
real $SL(2,R)$ spinors, 
$$
(Q^1_+, Q^2_-) \equiv Q_\alpha~, \quad (-Q^2_+, Q^1_-) \equiv S_\alpha~, 
$$
we can rewrite \p{A23c} in this particular case as 
\bea  \lb{D3N1}
&&\{ Q_\alpha, Q_\beta\} = \{ S_\alpha, S_\beta\} = 2P_{\alpha\beta}~, 
\; P_{11} = P_\pp~, \; P_{22} = P_=~, \; P_{12} = P_{21} = Z_2~, \nn \\ 
&& \{ Q_\alpha, S_\beta\} =0~.
\eea
Thus \p{B6} can be interpreted as a static-gauge form of the action for an
$N=2$, $D=3$ superstring in the formulation with manifest world-sheet 
supersymmetry. It could be recovered by the world-volume dimensional 
reduction from the PBGS $D=3$ action describing the $N=1,\; D=4$ 
supermembrane \cite{IK}\footnote{The general case \p{Z1cond} amounts to 
an extension of \p{D3N1} by tensorial central charges which is 
a reduction of the tensorial-charge extension of $N=1, D=4$ 
Poincar\'e superalgebra \cite{FeP}-\cite{GGHT}.}.          

Similarly to the $N=(2,0) \rightarrow N=(1,0)$ breaking case (Subect.3.2),
the scalar Goldstone superfield $\pi$ can be replaced by the covariant
superfield strength $\pi_v$ of the $N=(1,1)$ vector multiplet \cite{HPS} 
\be
\pi\Rightarrow \pi_v=\Dpo V_- +\Dmo V_+~.\lb{B8}
\ee
Here, the real fermionic superfields $V_\pm$ are the $N=(1,1)$ gauge
potentials, 
$$
\delta V_\pm = D^1_\pm \Lambda~. 
$$
Like in the case considered in Sect. 3.2,  $\pi_v$ in the Wess-Zumino
gauge differs from $\pi$ \p{compxi} merely by the replacement
\be
F \;\;\Rightarrow \;\; F_v = \Pp\, v_= - \Pm v_\pp~, \lb{strvect}   
\ee
where $ (v_=(x),\;v_\pp\,(x))$ is a $D=2$ abelian gauge field. All the
above
transformation properties and formulas remain valid. The bosonic action
\p{boson21}, with $F_v$ substituted for the auxiliary field $F$ is the $D=2$
Dirac-Born-Infeld (DBI) action:
\be
S_1^{bos}(v)  = {1\over 2f^2} \int d^2x \left[1 -
\sqrt{1 -4(\Pp\,p\Pm p + F_v^2)}\right]~. \lb{BI21}
\ee 
So in this case the superfield action \p{B6} can be interpreted as 
a manifestly world-sheet supersymmetric form of the action of
a D1-brane in $D=3$. It can equally be reproduced by a world-volume
dimensional reduction of the space-filling D2-brane \cite{IK}. 

Note that the nonlinear realization of partial breaking  $N=(2,2)
\rightarrow N=(1,1)$ with the Goldstone $N=(1,1)$ gauge multiplet can
also be derived from the same $N=(2,2)$ superfield transformation law
\p{fintran}
as the starting point, but with $\hat\phi$
replaced by the chiral superfield strength of $N=(2,2)$ vector multiplet,
\be
\hat\phi \;\rightarrow \; \hat\phi_v = \bar D_+\bar D_- {\cal V}~.
\lb{hatphiv}
\ee
Here, ${\cal V}$ is a real gauge prepotential \cite{HPS,RV},
\be
\delta {\cal V} = \lambda + \bar\lambda~, \quad D_+\lambda = 
\bar D_-\lambda =0~.  
\ee
The transformation \p{fintran} for $\hat\phi_v$ is induced by the
following modified transformation of ${\cal V}$, 
$$
\delta {\cal V} = -\left[(\epsilon^+_2\theta^- -
\epsilon^-_2\theta^+)\bar\theta^-\bar\theta^+ + \mbox{h.c.} \right] +
(\epsilon Q){\cal V}~.
$$ 
While written through $\phi_v(\pi_v)$, the invariant \p{B3d} is recognized as 
a generalized Fayet-Iliopoulos term, with the parameter $b$ being the 
$\theta$-angle (cf. the discussion in Sect. 3.2).

As the classical Born-Infeld dynamics is trivial in 
$D=2$, one can expect the above  D1-model to be related to the previous
$N=2, D=3$ superstring model with the world-sheet scalar multiplet 
$\pi(x,\theta^\pm)$. Let us elaborate on this relationship at the 
level of bosonic actions. The equations for the gauge field in \p{BI21} yield
\be      
F_v = \gamma \, \sqrt{1 -4(\Pp\,p\Pm p + F_v^2)}~, 
\ee
where $\gamma $ is an arbitrary integration constant. Solving
this for $F_v$ and substituting the result back into \p{BI21} 
brings the latter to the form    
\be
S_1^{bos}(v)  = {1\over 2\tilde{f}^2} \int d^2x \left[1 -
\sqrt{1 -4\Pp\,p\Pm p} + \frac{4\gamma^2}{1 + \sqrt{1+4\gamma^2}}\right]~.
\lb{BI212}
\ee
This coincides with the Nambu-Goto action up to a shift by a 
positive constant. This shift is the net effect of the presence 
of the $D=2$ gauge fields in the initial DBI action \p{BI21} (at the 
classical level). Precisely the same bosonic action can 
be recovered after elimination of the auxiliary field $F$ in the 
modified $D=3$ superstring PBGS action \p{B66} with $b = -\gamma$. Thus we 
conclude that the $D=3$ D1-brane world-sheet effective action $S_1(v)$ is
classically equivalent on shell to the modified $N=2$, $D=3$ superstring
PBGS action \p{B66}. Adding the topological $\theta$-term 
$\sim {\theta\over 2\pi} F_v$ to \p{BI21} does not affect this conclusion. 

As a last remark we note that the $\epsilon_2$ transformations of the
gauge potentials, which  produce for $\pi_v$ \p{B8}  
the nonlinear-realization transformation law \p{vreal}, are as follows
\bea
\delta V_+=i\emt\tpo\tmo -\emt \sigma(\pi_v)~, \quad
\delta V_-=-i\ept\tpo\tmo -\ept \sigma(\pi_v)~.\lb{B9}
\eea
It is easy to check that the algebra of $N=(2,2)$ supersymmetry 
is essentially modified on the gauge potentials $V_\pm$ as compared to the
gauge-invariant field strength $\pi_v$. An analogous modification of 
the algebra of spontaneously broken supersymmetry on gauge potentials was
earlier found for the D3-brane \cite{BG,IZ} and D2-brane \cite{Z3} in the
PBGS formulations.

\setcounter{equation}{0}
\section{\lb{F} Partial breaking N=(2,2) to N=(2,0)}

The partial spontaneous breaking $N=(2,2) \rightarrow N=(2,0)$ has been
considered earlier  in ref. \cite{HP}. We shall discuss this case
as an illustration of the general methods.

Our starting point will be again the chiral $N=(2,2)$ superfield
$\phi$ with the generic transformation properties \p{tranphi}, \p{explicQ}. 
This time $\phi$ is required to obey a sort of the holomorphy condition with 
respect to the central charges
\be
(Z_1 - iZ_2)\phi = 0~. \lb{holom}
\ee     
Then, by means of the appropriate frame rotation, the generators
\p{explicQ} can be brought into the form 
\bea 
&&\hat{Q}_+ = Q_+~, \quad \hat{\bar Q}_+ = \bar Q_+~, \nn \\
&&\hat{Q}_- = Q_- - 2\theta^+(Z_3 +i Z_4)~,\quad \hat{\bar Q}_- = \bar Q_-
+ 2\theta^+ (Z_1 +iZ_2)~.
\lb{Q2220} 
\eea
The $N=(2,0)$ generators $\hat{Q}_+~, \;\hat{\bar Q}_+$ contain no central
charge terms and hence correspond to unbroken supersymmetry. For the  
$N=(0,2)$ part of $N=(2,2)$ supersymmetry to be fully
broken, we have two obvious alternatives
\be
(a)\; (Z_1 + iZ_2)\phi = 0~, \;\; (Z_3 + iZ_4)\phi \neq 0; 
\qquad (b)\; (Z_1 + iZ_2)\phi \neq 0~, \;\;  (Z_3 + iZ_4)\phi = 0~. 
\lb{altern} 
\ee
Note that in the case of simultaneous presence of both $Z_1 + iZ_2$ and  
$Z_3 + iZ_4$ acting as shifts of $\phi$ with the relative coefficient 
being a pure phase, one can always find a real
combination of the generators $\hat{Q}_-~, \;\hat{\bar Q}_-$ containing 
no central charges. In this case we are facing the $3/4$ partial
breaking option, with only one real $N=(0,1)$ supersymmetry being 
broken. This interesting option deserves a special analysis which is 
however beyond the scope of the present paper.\footnote{Note that 
this possibility arises only if two sorts of complex central charge 
generators are simultaneously present in the $N=(2,2)$
superalgebra \p{A8}, \p{A23c}. This is in agreement with the general
conclusions of refs. \cite{FeP}-\cite{GGHT} that such an option in 
the case of $N=1, D=4$ supersymmetry is possible only in the presence 
of tensorial central charges (see the footnote after eq. \p{q1alg2}).} 

If the relative coefficient is not a 
pure phase, $N=(0,2)$ supersymmetry is fully broken, i.e. we are facing 
the $1/2$ breaking of $N=(2,2)$. We shall not consider this most general
mixed case of the $N=(2,2) \rightarrow N=(2,0)$ partial breaking, but limit our
study to the above two extremal cases for simplicity.    

These two versions give rise to two different $\epsilon^-$
transformation laws for $\phi$
\bea 
&&(a)\;\delta \phi =\emi\,\tp +(\emi Q_-+\bem
\bar{Q}_-)\phi~, \;\; (Z_3 + iZ_4)\phi = -{1\over 2} \lb{F22} \\ 
&&(b)\; \delta \phi =\bem\,\tp +(\emi Q_-+\bem
\bar{Q}_-)\phi~, \;\; (Z_1 +iZ_2)\phi = {1\over 2}~.\lb{F22b}
\eea 
In both of them two $\epsilon^-$ supersymmetries are broken with 
$\xi^- \equiv - D_+ \phi$ as the corresponding Goldstone fermion
\be
(a)\;\delta \xi^- = 
\emi + (\emi Q_-+\bem
\bar{Q}_-)\xi^-~, \;\; (b) \;\delta \xi^- = 
\bem + (\emi Q_-+\bem
\bar{Q}_-)\xi^-~.
\ee
The central charges shift $\phi$ by a complex parameter. 
Like in the previously considered examples, the specific realization 
of the central charges on $\phi$ in \p{F22}, \p{F22b} was chosen 
for convenience: actually, the coefficient before the inhomogeneous 
pieces in these transformation laws are defined up to an arbitrary
(in general, complex) rescaling of $\phi$. The first version is easier to
treat by our procedure, so below we just specialize to it. 

As in the previously studied cases, it is convenient to deal  
with the superfields of unbroken supersymmetry, in the present case 
with $N=(2,0)$ superfields. The $\tm$ expansion of the $N=(2,2)$ chiral
superfield $\phi$ reads  
\be
\phi(\xpl,\xml,\tp,\tm)=(1-i\tm\btm\Pm )u(\xpl,\xm,\tp)+
\tm\eta^+(\xpl,\xm,\tp)~,\lb{F1}
\ee
where $u$ and $\eta^+ $ are the $N=(2,0)$ chiral superfields, 
respectively bosonic and fermionic, 
\be
\bar D_+u = \bar D_+ \eta^+ = 0~. \lb{chir20}
\ee
The $N=(2,2)$ superfield transformation \p{F22} induces the following 
ones for these $N=(2,0)$ superfield components,  
\be
\delta u = \emi (\tp +\eta^+)~, \quad \delta \eta^+ = -2i\bem \Pm u~. 
\lb{20tran}
\ee

Let us now construct the corresponding nonlinear realization by the 
universal procedure employed in the previous cases.

We perform finite $\epsilon^-$ transformations of the superfields 
$\eta^+$ and $D_+u$, then change, in the transformed superfields,  
the Grassmann parameters $\epsilon^-, \bar\epsilon^{\;-}$ to $-\Psi^-,\;
-\bar\Psi^-$ where $\Psi^-, \bar\Psi^-$ are the Goldstone $N=(2,0)$ 
fermionic superfields with the universal $\epsilon^-$ transformation law 
\be
\delta \Psi^-=\emi -i(\emi\bar\Psi^- +\bem\Psi^-)\Pm \Psi^-~, \quad 
\delta \bar\Psi^- = \overline{\delta \Psi^-}~, \lb{F6b}
\ee
and, finally, equate to zero the resulting ``tilded'' superfields with 
the homogeneous nonlinear transformation law under the $\epsilon^-$
transformations. In this way we arrive at the following 
set of covariant equations  
\bea
&& \Dp u + \Psi^-(1+ \Dp\eta^+) -i \Psi^-\bar\Psi^-\Dp\Pm u=0~,
\nn\\
&& \bDp \bar{u} + \bar\Psi^-(1- \bDp\bar\eta^+)+i \Psi^-\bar\Psi^-
\bDp\Pm u=0~,
\lb{F5}\\
&&\eta^+ + 2i\bar\Psi^-\Pm u + i\Psi^-\bar\Psi^-\Pm\eta^+=0~,\nn\\
&&\bar\eta^+ -2i\Psi^-\Pm \bar{u} - i\Psi^-\bar\Psi^-\Pm
\bar\eta^+=0~.\lb{F5b}
\eea
As in the previous cases, the first two equations set the equivalence 
relation between the linear- and nonlinear-realization Goldstone fermion 
superfields, while the last pair of equations (together with the first one)
serves to 
covariantly express $\eta^+, \bar\eta^+$ in terms of the irreducible
Goldstone chiral $N=(2,0)$ superfield $u$.

These equations can be solved explicitly. As the first step, from
\p{F5} and \p{F5b} we find the following relations, 
\be
\eta^+   = \frac{2i\, \bDp{\bar u}\Pm u }{ 1-\bDp{\bar\eta}^+ }\; , \quad
{\bar\eta}^+ =  -\frac{2i\,\Dp u\Pm {\bar u}}{1+\Dp{\eta}^+}\;.\label{F7}
\ee
After some algebra, the relations \p{F7} can be rewritten as
\bea
&& \eta^+ = 2\bDp\left[i\bar{u}\Pm u +  \frac{2\,\Dp u\bDp \bar u\Pm
u\Pm
\bar{u}}{(1+\Dp \eta^+)(1-\bDp \bar\eta^+)}\right]~, \nn \\
&& \bar\eta^+ = -2 \Dp\left[i u\Pm {\bar u} -
  \frac{2\,\Dp u\bDp \bar u\Pm u\Pm
\bar{u}}{(1+\Dp \eta^+)(1-\bDp \bar\eta^+)}\right]~.
\lb{F7b}
\eea
The numerators of the second terms in the square brackets 
already contain the maximal number of fermions. Therefore we have 
to find only the ``effective'' expressions for the denominators, including 
no fermions without derivatives. Hitting \p{F7b} by the corresponding 
spinor derivatives and discarding all the terms in which  fermions appear 
without $x$-derivatives on them, one gets the following system of 
equations:
\be\label{addon}
\left( \bDp\bar\eta^+\right)_{eff} -
\frac{B}{1+\left(\Dp \eta^+\right)_{eff} }=0\;
,\quad
\left( \Dp\eta^+\right)_{eff}+
\frac{\bar B}{1 -\left(\bDp \bar\eta^+\right)_{eff} }=0\; ,
\ee
where
\be
B= 4\,\Pp\, u\Pm{\bar u}~,\qq \bar{B}= 4\,\Pp\, {\bar u}\Pm u~.\lb{F9}
\ee
These algebraic equations have the following solution (with $\eta^+$ 
admitting a power expansion around the point $u=0, \bar u =0$) 
\bea
&& \left(\Dp \eta^+ \right)_{eff} = -{1\over 2}\left(1 - B + \bar B - 
\sqrt{(1-B-\bar{B})^2-4B\bar{B}} \right)~, \nn \\ 
&& \left(\bDp \bar\eta^+ \right)_{eff} = {1\over 2}\left(1 + B - \bar B - 
\sqrt{(1-B-\bar{B})^2-4B\bar{B}} \right)~.
\eea
Substituting this into \p{F7b} we get the final answer for $\eta^+$ in
terms of $u, \bar u$ 
\be
\eta^+(u, \bar u) = 2\bDp\left[i\bar{u}\Pm u +\frac{4\,\Dp u\bDp {\bar
u}\Pm u\Pm
\bar{u}}
{1-B-\bar{B}+\sqrt{(1-B-\bar{B})^2-4B\bar{B}}}\right]~,\lb{F8}
\ee
It is straightforward, though tedious, to check that 
the $\epsilon^-, \bar\epsilon^{\;-}$ transformations of this $\eta^+(u,
\bar u)$ are in accordance with the law \p{20tran}. Its chirality is
manifest in the notation \p{F8}.

To construct the invariant action, we again first observe that the
basic $N=(2,2)$ superfield $\phi$ with the transformation law \p{F22}
admits two invariants \p{B3e} and \p{B3d}. In terms of the $N=(2,0)$
superfields defined in \p{F1} these invariants take the form
\bea
&&S_1= {1\over f^2}\int d^6z\,\phi\bar\phi= {1\over f^2}\int
d^4z^{(+)}\left[i(u \Pm\bar
u - \bar u \Pm u) +
\eta^+\bar\eta^+ \right]~,\lb{Acti1}\\ 
&& S_2 = \frac{1}{2 f^2}\int d^4\zeta_L\;\phi
+\mbox{c.c.}= -{1\over 2 f^2}\int d^2x D_+ \eta^+ + 
\mbox{c.c.}~.\lb{Acti2}
\eea

In full analogy with the previous cases, these invariants prove to be  
identical to each other for the nonlinear realization constructed. 
Substituting, e.g., the expression \p{F8} into $S_2$,  we obtain 
\be
S_2(u) ={1\over f^2}\int d^4z^{(+)}\left[ i(u\Pm \bar u - \bar u \Pm u) 
- \frac{8\, \Dp u \bDp \bar u\Pm u\Pm \bar{u}}
{1-B-\bar{B}+\sqrt{(1-B-\bar{B})^2-4B\bar{B}}}\right].\lb{F10}
\ee
Using eqs \p{F7} - \p{F8}, it is easy to show that 
\be
S_2(u) = S_1(u)~. \lb{F100}
\ee   

The bosonic part of the action \p{F10} reads 
\be
S^{bos}(u) = {1\over 2f^2}\int d^2x \left\{1 - 
\sqrt{(1-B-\bar{B})^2-4B\bar{B}}\right\} \lb{BI220} 
\ee
which is the static-gauge form of the Nambu-Goto action of the string in  
$D=4$ Minkowski space, the real and imaginary parts of $u(x)$ being two
transversal coordinates. The authors of \cite{HP} have explicitly shown
that \p{F10} is the manifestly world-sheet supersymmetric form of the
Green-Schwarz action of an $N=1$ superstring in $D=4$ (they used 
a different parametrization for the Goldstone multiplet, representing 
it by a superfield subject to some nonlinear chirality constraints, 
in contrast to the ordinary chiral Goldstone superfield $u$ 
in our approach). This form 
arises after fixing local fermionic $\kappa$-symmetry so as to kill 
half of the target spinor coordinates. It is straightforward to see 
that the algebra \p{A23c} with $Z_1 =Z_2=0$ coincides with the 
$N=1, D=4$ Poincar\'e superalgebra after combining the $D=2$ spinor 
generators into $D=4$ Weyl spinors via
$$
(\bar Q_+,\; Q_-) \equiv Q_\alpha ~, \quad (Q_+,\; \bar Q_-) \equiv \bar
Q_{\dot\alpha}   
$$
and identifying $P_\pp~,\; P_= $ and $Z_3 + iZ_4$ with the components 
of the 4-momentum generator. Note that the action \p{F10} can be also
recovered by world-volume dimensional reduction from the $D=4$ chiral 
Goldstone superfield action of \cite{BG2} which corresponds to the $N=1$ 
super 3-brane in $D=6$.   

We conclude by a few comments. 

First, in the case under consideration, without using derivatives, 
it is impossible to construct a homogeneously transforming 
superfield like \p{B0b} because the central charge acts 
as a complex shift of $\phi$. Hence the above nonlinear 
realization, as opposed to the one considered in the previous Section, cannot be 
reproduced alternatively by imposing a nilpotency condition on some proper 
superfield along the lines of ref. \cite {RT,GPR}. At the same time, our 
general procedure works pretty well in this case, too. It definitely
amounts to imposing some covariant constraints on the linear-realization 
Goldstone superfield $\phi$, but they necessarily include derivatives and 
it is rather hard to immediately guess their precise form.  

Second, the Goldstone chiral $N=(2,0)$ superfield 
$u$ collects only physical degrees of freedom (transverse string coordinates, or Goldstone 
fields for the complex central charge,  and the complex Goldstone fermion). 
So the case under consideration cannot be related to a D1-brane, 
in contrast to the option considered in  Sect. 4. It seems also impossible
to make use of the vector $N=(2,0)$ multiplet as an alternative Goldstone
one to represent the partial breaking pattern considered here. Indeed, this
multiplet can be described by the chiral fermionic superfield strength
$\Xi^+_v$, eq. \p{20str}. It has the same $SO(1,1)$ weight as 
$\theta^+$ and for this reason cannot be utilized as a Goldstone fermion 
for the partial breaking $N=(2,2) \rightarrow N=(2,0)$ (such a fermion 
should have the same weight as $\theta^-$). On the other hand, it 
could support the $1/2$ breaking $N=(4,0) \rightarrow N=(2,0)$. A
difference from the case $N=(2,0) \rightarrow N=(1,0)$ with the vector 
Goldstone $N=(1,0)$ multiplet (Sect. 3.2) is that the $N=(2,0)$ vector
multiplet contains a scalar auxiliary field (in addition to the complex
chiral fermion and $D=2$ gauge field strength). So the corresponding 
Goldstone superfield action is expected to be less trivial. 

Third, let us note that the initial linear realization \p{Q2220} 
(with $Z_1 +iZ_2 = 0$) is equivalent, by the $Z$-frame rotation 
with $A=2\theta^-\theta^+$, to the following one
\bea
&&\hat Q_+ = Q_+ - 2\theta^-(Z_3 +iZ_4)~, \quad  
\hat{\bar Q}_+ = \bar Q_+ \nn \\
&& \hat Q_- = Q_-~, \quad   \hat{\bar Q}_- = \bar Q_-~. 
\eea
This choice corresponds to the PBGS pattern $N=(2,2) \rightarrow N=(0,2)$. 
The relevant nonlinear-realization Goldstone multiplet is represented 
by a chiral $N=(0,2)$ superfield $u(\xp, \xml, \theta^-)$. The Goldstone 
superfield action is related to \p{F10} by a mirror symmetry and 
amounts to an alternative gauge-fixing of $\kappa$-symmetry in the 
$N=1, D=4$ superstring Green-Schwarz action.
 
\setcounter{equation}{0}
\section{\lb{D} Partial breaking N=(2,2) to N=(1,0)}
As was noticed in Sect.4, before imposing any restrictions on 
the central charges in \p{explicQ} (apart from the condition \p{chirint} 
required for the existence of chiral $N=(2,2)$ superfields), 
one can define a chiral superfield with the transformation law 
corresponding to $1/4$ partial breaking of $N=(2,2)$ supersymmetry, 
i.e., such that only one real supersymmetry acts by homogeneous 
transformations. By means of the appropriate $Z$-frame rotation, 
the generators \p{explicQ} can be brought into the form 
\bea \lb{14pbgs}
\hat{Q}_+ &=& Q_+ - 2\theta^- (Z_1 -iZ_2)~, 
\nn \\ 
\hat{\bar Q}_+ &=& \bar Q_+ + 2\theta^- (Z_1 -iZ_2)~, \nn \\
\hat{Q}_- &=& Q_- - 2\theta^+[(Z_3 +i Z_4)-(Z_1-iZ_2)]~,\nn \\
\hat{\bar Q}_- &=& \bar Q_- + 2\theta^+ (Z_1 +iZ_2)~.
\eea     
With all the central charge generators realized as complex shifts of the 
chiral superfield $\phi$, it follows from the general transformation law, 
\be \lb{14transf}
\delta \phi = (\epsilon \hat Q)\phi~,
\ee
that only the $\epsilon^+_1$ supersymmetry generated by 
$\hat{Q}^1_+ \sim \hat{Q}_+ + \hat{\bar Q}_+$ is unbroken, because 
only its transformation contains no central charge terms. 

Like in the previous examples, there exist quite a few possibilities 
to choose a particular realization of central charges with preserving 
the above crucial property. In this paper we do not aim at the exhaustive 
analysis of all possibilities, and choose this realization so as to make 
the computations most feasible,   
\be
(a)\; (Z_1 +iZ_2) \phi = 0~, \quad (b)\; (Z_1 -iZ_2)\phi = {i\over 2}~, 
\quad \;(c)\;  (Z_3 +iZ_4)\phi =-(Z_1 -iZ_2)\phi=  -{i\over 2}~. \label{cz3} 
\ee
With this ansatz, the transformation law \p{14transf} becomes 
\be
\delta \phi = -2i\theta^+ \epsilon^- -\sqrt{2}\theta^-\epsilon^+_2 +
(\epsilon Q)\phi~.   \label{14tr22}
\ee

Let us now consider the $N=(1,0)$ decomposition of the $N=(2,2)$ chiral
superfield
\be
\phi=(1+i\tpt\Dpo)[(1-\tmo\tmt\Pm) ( \phi_1+i\phi_2)(x,\tpo)+
(\tmo+i\tmt)(\pi^++i\eta^+)(x,\tpo)]~,\lb{D0}
\ee
where $\phi_1,~\phi_2$ and $\pi^+,~\eta^+$ are the real bosonic and 
fermionic $N=(1,0)$ superfields. 
The $N=(2,2)$ superfield transformation law \p{14tr22} implies 
the following transformation laws for these $N=(1,0)$ projections
\bea
&&\delta \phi_1= i\epsilon_1^- \left( \tp + \eta^+\right) -\epsilon_2^+ D_+
\phi_2 + i\epsilon_2^- \pi^+~,\nn\\
&&\delta \phi_2= i\epsilon_2^-\left( \tp+\eta^+\right)+\epsilon_2^+
D_+\phi_1- i\epsilon_1^- \pi^+~,\nn\\
&&\delta\pi^+= -\epsilon_2^+\left( 1+ D_+\eta^+\right) 
  +\epsilon_1^-\Pm\phi_2-\epsilon_2^-\Pm\phi_1,\nn\\
&&\delta\eta^+= \epsilon_2^+ D_+\pi^+ -\epsilon_1^-\Pm\phi_1-
  \epsilon_2^-\Pm \phi_2 \label{22phi}
\eea
(hereafter we omit the index 1 of the real $N=(1,0)$ superspace 
Grassmann coordinate, $\theta^+_1 \rightarrow \theta^+$). 
As is seen from \p{22phi}, the spinor superfield $\pi^+$ 
is the Goldstone fermion for the $N=(0,1)$ supersymmetry with 
the parameter $\epsilon^+_2$. Two other Goldstone fermions which 
are required by partial breaking of the remaining $N=(0,2)$ supersymmetry 
are the spinor derivatives of the superfields $\phi_1$ and $\phi_2$: 
\bea
&& \xi^1_+ \equiv iD_+\phi_1 \; \Rightarrow \; 
\delta \xi^1_+ = \epsilon_1^-\left( 1+D_+\eta^+ \right)+
 \epsilon_2^+ D_+ \xi^2_+ 
          + \epsilon_2^-D_+\pi^+ \;, \nn \\
&&\xi^2_+ \equiv i D_+\phi_2 \;  \Rightarrow \;
\delta \xi^2_+ = \epsilon_2^-\left( 1+D_+\eta^+ \right) -
  \epsilon_2^+ D_+ \xi^1_+  - \epsilon_1^-D_+\pi^+ \;. \label{22xi}
\eea
The superfields $\phi_1$ and $\phi_2$ are the Goldstone ones for 
two independent central-charge shifts of $\phi(\zeta_L)$.

The superfield $\eta^+$ transforms homogeneously with respect to
all three spontaneously broken supersymmetries. In virtue of the
transformation properties \p{22phi} one can construct
the simplest off-shell invariant
\begin{equation}\label{22action1}
S= -i\int d^3 z^{(+)} \eta^+ = \int d^2 x D_+\eta^+ 
\end{equation}
(hereafter, for convenience, we put normalization factors before actions 
equal to 1). It is the $N=(1,0)$ superfield form of the universal action  
$S_2$ defined in \p{B3d}, which, along with another such action, $S_1$, 
eq. \p{B3e}, are invariant under the realization \p{14tr22}, 
similarly to the cases of other inhomogeneous realizations of $N=(2,2)$ 
supersymmetry considered in the previous Sections. The precise form 
of these two invariants is as follows 
\bea
&& S_1 \sim {1\over 4} \int d^6 z \,\phi\bar\phi = 
\int d^3z^{(+)}\left[\sum_{i=1,2}\Pm\phi_i D_+\phi_i -
i\left(\pi^+ D_+\pi^+  + \eta^+ D_+\eta^+ \right)\right]~, \lb{14inv1} \\
&& S_2 \sim {i\over 4}\int d^4\zeta_L\, \phi +\mbox{c.c.}  
= -i\int d^3 z^{(+)}\,\eta^+~. \lb{14inv2}   
\eea

Like in the previous cases, one possibility 
to make the invariant \p{22action1} meaningful is to covariantly express
the superfield $\eta^+$ in terms
of Goldstone superfields $\xi_+^{1,2}, \pi^+$. In other words, we
have to construct from Goldstone superfields the spinor superfield
which transforms as $\eta^+$ with respect to the full $N=(2,2)$
supersymmetry. The idea of such a construction is the same as before: 
we write the finite supersymmetric transformation of our superfields 
$\xi_+^{1,2}, \pi^+,\eta^+$ and replace the parameters 
$\left\{ -\epsilon_2^+, -\epsilon_{1,2}^+\right\}$  by the Goldstone
fermionic superfields of nonlinear realizations 
$\left\{ \psi_2^+,\psi_{1,2}^+\right\}$ :
\bea
&&{\tilde\xi}^1_+= \xi^1_+ 
 -\psi_2^+ D_+ \xi^2_+ -
  \psi_1^-\left( 1+D_+\eta^+ \right)
          - \psi_2^-D_+\pi^+ \ldots\;, \nn \\
&& {\tilde\xi}^2_+=  \xi^2_+ +
  \psi_2^+ D_+ \xi^1_+  + \psi_1^-D_+\pi^+
-\psi_2^-\left( 1+D_+\eta^+ \right) +\ldots \;, \nn \\
&&\tilde\pi^+ = \pi^+ +\psi_2^+\left( 1+ D_+\eta^+\right) 
  -\psi_1^-\Pm\phi_2 +\psi_2^-\Pm\phi_1+\ldots \;,\nn\\
&&\tilde\eta^+= \eta^+ -\psi_2^+ D_+\pi^+ +
 \psi_1^-\Pm\phi_1+
  \psi_2^-\Pm \phi_2+\ldots \; .\label{22eq}
\eea
Here we have explicitly written down only linear terms, denoting the
higher-order terms by dots. The transformation properties of the 
universal Goldstone fermionic superfields are analogous to those 
given earlier, so we do not explicitly quote them. 

Once again, one can check that the new superfields  
$\tilde{\xi}_+^{1,2}, \tilde{\pi}^+,\tilde{\eta}^+$ transform homogeneously 
and hence can be covariantly equated to zero. These 
conditions provide the sought system of equations relating
Goldstone fermions of the linear $\left\{ \xi_+^{1,2}, \pi^+\right\}$
and nonlinear $\left\{ \psi_2^+,\psi^-_{1,2}\right\}$
realizations and also serving to covariantly express 
the superfield $\eta^+$ in terms of the remaining superfields.

The system \p{22eq} is much more complicated than those we encountered
in the previous Sections. It can be solved by iterations, but the solution 
looks not too enlightening. Let us concentrate on the pure bosonic part 
of the action which we firstly choose to coincide with the simplest 
invariant \p{22action1}. In the bosonic limit we can keep in eqs. \p{22eq} 
only the terms written explicitly, because the rest is at least
of the third order in fermions. Therefore, in this approximation 
our systems can be written as 
\bea
&&\left( \begin{array}{c} \pi^+ \\ \xi^1_+ \\ \xi^2_+ \end{array}
\right) - A \left( \begin{array}{c} 
 -\psi_2^+ \\ \psi^1_- \\ \psi^2_- \end{array}
\right)=0 \; , \label{22eq1} \\
&& \eta^+ -\psi_2^+ D_+\pi^+ +
 \psi_1^-\Pm\phi_1+
  \psi_2^-\Pm \phi_2 = 0 \; , \label{22eq2}
\eea
where
\be
A= \left( \begin{array}{ccc} 1 +D_+\eta^+ & \Pm\phi_2 &-\Pm\phi_1 \\
   -D_+\xi^2_+ & 1+D_+\eta^+ &  D_+ \pi^+ \\
  D_+ \xi^1_+ & -D_+\pi^+ & 1 +D_+\eta^+ \end{array} \right) \;.
\ee
Now we can solve the equation \p{22eq1} for  
$\left\{ \psi_2^+,\psi^-_{1,2}\right\}$ and substitute
the solution into \p{22eq}. After hitting \p{22eq2} with one spinor
derivative we finally obtain the equation
\be\label{22eq3}
D_+\eta^+ =- \left( D_+\pi^+ ,\; \Pm\phi_1, \; \Pm\phi_2\right)
A^{-1} \left( \begin{array}{c} D_+\pi^+ \\ D_+\xi^1_+ \\ D_+ \xi^2_+ 
        \end{array} \right) \; .
\ee
Let us recall that $a\equiv D_+\eta^+ $ is the bosonic Lagrangian density we
are looking for (see eq. \p{22action1}). Thus the bosonic Lagrangian 
can be found as a solution of eq. \p{22eq3} which amounts to 
the following quartic equation:
\be\label{22finaleq}
\left( a^2+2a+1+y\right)\left( a^2+a+y\right) +z\left( z+k\right) =0 \;,
\ee
where
\bea
&& k\equiv D_+\pi^+\;,\; y= k^2+\Pm\phi_1\Pp\phi_1+\Pm\phi_2\Pp\phi_2 \;, \nn\\
&& z=\Pm\phi_1\Pp\phi_2 - \Pm\phi_2\Pp\phi_1 \; .
\eea
To find the solution of eq.\p{22finaleq} to low orders in the fields, 
we firstly employ in \p{22finaleq} the equation of motion for the 
auxiliary field $k$ :
\be\label{22eqk}
\frac{\partial}{\partial k} a = 0\;\Rightarrow\; 
2k\left( 2a^2 +3a+2y+1\right)+z=0 \;.
\ee
The solution of the system \p{22finaleq} plus \p{22eqk}, 
up to eighth order in physical fields, reads
\bea\label{22pert}
&&a=-{\tilde y}-\frac{3}{4}z^2-{\tilde y}^2 -2{\tilde y}z^2-2{\tilde y}^3 -
  \frac{19}{16}z^4 -\frac{25}{4}{\tilde y}^2z^2 -5{\tilde y}^4 +\ldots \;,\\
&& k=-\frac{z}{2}\left( 1+{\tilde y}+\frac{7}{4}z^2+2{\tilde y}^2+
  \frac{11}{2}{\tilde y}z^2 +5{\tilde y}^3\right) +\ldots \; ,
\eea
where
$$  {\tilde y}= \Pm\phi_1\Pp\phi_1+\Pm\phi_2\Pp\phi_2 \;.$$
One can find a general solution of eq. \p{22finaleq}, but even after
eliminating the auxiliary field $k=D_+\pi^+$ this solution does not look
very informative.

To find a concise expression for the bosonic action, let us 
recall that, besides the Lagrangian $L_1=a$, we have another one,
which is equal to the free Lagrangian of our original linear $N=(2,2)$ 
supermultiplet
\be
L_{free}=a^2+y \;
\ee
(this expression is obtained from \p{14inv1} after integrating over $\theta^+$
and discarding the fermions). Let us define a new bosonic Lagrangian 
as some combination of $L_{free}$ and $L_1$ (up to an overall normalization 
factor),
\be\label{lagrfin}
L = L_{free}+\alpha L_1 \equiv a^2 +\alpha a +y \:,
\ee
where $\alpha$ is an arbitrary parameter for the time being.

The field $k$ is auxiliary, and therefore it can be eliminated by solving 
its algebraic equation of motion:
\be\label{eq2}
\frac{\partial}{\partial k} L \equiv
(2a+\alpha)\frac{\partial }{\partial k} a
+2k =0 \Rightarrow   \frac{\partial }{\partial k} a =
-\frac{2k}{2a+\alpha} \;.
\ee
Differentiating \p{22finaleq} with respect to $k$ and taking account 
of \p{eq2}, we get the equation 
\be\label{eq3}
-2k\left[ (2-\alpha)\left( L+(1-\alpha) a\right) +
 (1-\alpha )\left( L+(2-\alpha)a+1\right)\right] +(2a+\alpha )z =0 \; .
\ee
Now, fixing our free parameter $\alpha$ to 
\be
\alpha = 2~, \lb{fixalph}
\ee
we find the following unique solution of eqs. \p{22finaleq} 
and \p{eq3}, 
\be\label{sol}
\quad k=-z\; , \quad a^2+a+{\tilde y} =0 \:,
\ee
where
\be\label{tildey}
{\tilde y} = z^2 +\left( \Pp v\Pm v
+\Pm s\Pp s \right)|_{\theta=0} \:.
\ee
Therefore,
\be
a=-\frac{1}{2}\left(1-\sqrt{1- 4{\tilde y}}\right) \:,
\ee
and \p{lagrfin} becomes
\be\label{string}
L= -\frac{1}{2}\left(1-\sqrt{1-4{\tilde y}}\right) = a \; .
\ee
The physical Lagrangian $L_{phys} = -L$ describes a bosonic string 
in $D=4$ Minkowski space. The full Goldstone-superfield Lagrangian 
still drastically differs in fermionic terms from that of the 
standard $N=1, D=4$ superstring of Sect. 5 (three physical 
Goldstone fermions are present now as compared to two such 
fermions of the previous example).

We end with a few comments. 

First, recall that with the simultaneous presence of two sorts of 
central charges, $Z_1 -iZ_2$ and $Z_3 +iZ_4$, in the $N=(2,2)$ 
superalgebra \p{A23c}, the latter is a reduction of the 
tensor-central-charge-extended $N=1, D=4$ superalgebra \cite{FeP}-\cite{GGHT}. 
Since both sorts of central charges are present in the transformation 
law \p{14tr22} or \p{cz3}, from the $N=1, D=4$ Poincar\'e superalgebra 
standpoint the transversal coordinates $\phi_1$ and $\phi_2$ of the string 
are associated with some combinations of the standard four-momentum 
and the tensorial central charges. In other words, the ambient bosonic 
manifold of our system is non-trivially embedded into the product of 
ordinary Minkowski space and the space parametrized by coordinates conjugate 
to the tensorial central charges. A similar situation was observed in 
the superparticle models with $1/4$ breaking \cite{DIK}. 
A recent paper \cite{BVP} also discusses the possibility to
associate transverse brane coordinates with components of tensorial 
central charges, rather than with those of the conventional momentum operator. 
If one repeats for the present case
the model-independent analysis undertaken in ref. \cite{GGHT} 
to classify admissible BPS configurations in $N=1, D=4$ supersymmetry 
with tensorial central charges, one will find that a choice of central 
charges as in \p{cz3} allows indeed just for the $1/4$ breaking option. 
Yet, it is still unclear to us how the essentially on-shell analysis 
of \cite{GGHT} correlates with our approach which proceeds from off-shell 
superfield representations of $N=(2,2)$ supersymmetry. 

Second, the fractional patterns of PBGS other than $1/2$, in particular, 
the $1/4$ one, are known to naturally occur in systems of intersecting 
branes (see, e.g., \cite{OT}). It would be interesting to elaborate on a 
similar interpretation for our string-like $1/4$ system. A closely 
related problem is to construct an appropriate Green-Schwarz-type action. 
It should respect only one real $\kappa$-supersymmetry, break $D=4$ Lorentz 
invariance and, in the static gauge, reproduce the on-shell form of 
our Goldstone superfield action (still to be fully worked out).   

Third, like in the examples of Subsect.3.2 and Sect.4, one could 
pass to a ``$1/4$ super D1-brane'' by substituting the covariant 
field strength of the $D=2$ Maxwell field for the auxiliary field 
$k= D_+\pi^+$ in the above relations. As the auxiliary field now plays 
an important role in forming the correct string-type bosonic action, 
one expects this trick to have a greater impact on 
the structure of the relevant actions than compared to the 
previous cases. 

\setcounter{equation}{0}
\section{Conclusions}
In this paper we have constructed universal low-energy nonlinear Goldstone
superfield actions for a few patterns of partial breaking of $N=(1,1)$,
$N=(2,0)$ and $N=(2,2)$ supersymmetries in two dimensions, We have shown
that these actions provide a manifestly world-sheet supersymmetric 
description of some superstrings and D1-branes in flat $D=3$ and 
$D=4$ Minkowski backgrounds. One novelty of
our treatment is that we proceeded from a purely two-dimensional setting, 
without assuming embeddings into higher dimensions in advance. We found
that in most cases, for the partial breaking to occur, the supersymmetries
must necessarily be extended by appropriate central-charge generators. 
The latter produce pure shifts of scalar fields in the Goldstone
multiplets, allowing one to identify these fields with the transversal 
coordinates of some brane in the static gauge. 

As another novel point our 
construction systematically exploits the general relation between linear
and nonlinear realizations of spontaneously broken supersymmetries
\cite{IKa}.This allows us to deduce new equivalent forms of the Goldstone 
superfield actions in terms of the superfields of the full supersymmetry.
These superfields transform linearly (though inhomogeneously) under the 
broken part of the supersymmetry and are nonlinear functions of the basic 
irreducible Goldstone multiplets. In such a representation both kinds of
supersymmetries, broken as well as unbroken, are manifest. Besides 
the $1/2$ partial breaking options, we considered the $1/4$ breaking of
$N=(2,2)$ supersymmetry, showed the existence of a manifestly 
supersymmetric Goldstone superfield action for this PBGS pattern and found 
its bosonic piece. It would be of great interest to find the 
corresponding Green-Schwarz-type action (if existing).       

We did not aim to give an exhaustive analysis 
of all possible schemes and realizations of partial breaking of 
two-dimensional supersymmetries. Our goal was to construct 
low-energy Goldstone superfield actions for a few technically 
feasible cases. For the convenience of the reader, we summarize our 
basic examples in the Table below. We quote the superfields of 
linear realization (LR), the minimal Goldstone multiplets (GM), 
the non-vanishing central charges (CC), and the space-time interpretation of 
the Goldstone superfield actions.

\vspace{0.4cm}
\noindent\hskip 0.25cm $
\begin{array}{||c|c|c|c|c||}
\hline  \hline 
\mbox{PBGS} &\mbox{LR} & \mbox{GM}  & 
\mbox{CC}  & \mbox{Interpretation} \\ \hline\hline
(1,1)/(1,0) &\Phi(x,\theta^\pm_1) & u(x,\theta^+_1) &Z & 
N=1, D=3\, \mbox{sstr.}\\ \hline
 (2,0)/(1,0) &  \begin{array}{cc} \Xi^+(\xm, \xpl, \theta^+)\\
         \Xi^+_v(\xm, \xpl, \theta^+) \end{array} \hspace{-0.2cm}
 &\begin{array}{cc} \xi^+(x,\theta^+_1) \\
            \xi^+_v(x,\theta^+_1) \end{array}   \hspace{-0.2cm} 
&\begin{array}{cc}\mbox{No}  \\
          \mbox{No} \end{array}
&\hspace{-0.16cm}\begin{array}{cc} \mbox{left-chir. ferm.} \\
        \mbox{space-filling D1-br.} \end{array}\hspace{-0.25cm}
 \\  \hline
(2,2)/(1,1) & \begin{array}{cc}\phi(\zeta_L) \\\phi_v(\zeta_L) \end{array}
   & \begin{array}{cc} \pi(x, \theta^\pm_1) \\ \pi_v(x, \theta^\pm_1) \end{array}
   &\begin{array}{cc} Z_2 \\ Z_2 \end{array} 
   &\hspace{-0.16cm}\begin{array}{cc} N=2, D=3\, \mbox{sstr.} \\
      D=3\, \mbox{D1-br.} \end{array} \hspace{-0.25cm}\\ \hline
(2,2)/(2,0) &\phi(\zeta_L) 
&u(\xm, \xpl, \theta^+) &Z_3, Z_4 &N=1, D=4\, \mbox{sstr.} \\ \hline
(2,2)/(1,0) &\phi(\zeta_L) 
&\phi_{1,2}(x, \theta^+_1), 
\pi^+(x,\theta^+_1) &Z_1 - Z_3, Z_2 - Z_4 &? \\ 
\hline   \hline
\end{array}
$ 

\vspace{0.4cm}
We finish with a few comments and proposals for future study. 

When setting up linear realizations of the partially
broken supersymmetries, we proceeded from the simplest supermultiplets 
of the latter. In particular, all patterns of the $N=(2,2)$ 
supersymmetry breaking were realized on a single $N=(2,2)$ chiral
superfield, while the different PBGS options were realized by choosing 
different sets of central charges in the $N=(2,2)$ superalgebra. 
We could equally reproduce all these options, with the same 
final minimal Goldstone superfield actions, by starting from a
twisted-chiral $N=(2,2)$ superfield. In this case, the manifestly $N=(2,2)$
supersymmetric form of these actions is given by expressions like
\p{B3e} and \p{B3d}, with twisted-chiral superfields instead of the chiral
ones. These two equivalent representations of the same 
Goldstone superfield action are related to each other 
by mirror symmetry (see, e.g., \cite{MP}) which 
interchanges
chiral and twisted-chiral superfields and amounts to the twists $\theta^+ 
\leftrightarrow \bar\theta^+$ or $\theta^- \leftrightarrow \bar\theta^-$
accompanied by appropriate reflections of the irreducible Goldstone
superfields. One more possibility is to start from  a 
semi-chiral $N=2$ superfield \cite{BLR}. Once again, based on the
universality
of the Goldstone superfield actions, we expect the corresponding linear
realizations of the $N=(2,2)$ PBGS patterns to lead, upon applying our 
general procedure, to the same nonlinear realizations and minimal
Goldstone superfield actions modulo field redefinitions.  
     
New opportunities arise when more than one $N=(2,2)$ superfield is 
incorporated simultaneously. For instance, a chiral plus a twisted chiral
superfield form 
a simplest off-shell representation of $N=(4,4)$ supersymmetry, the twisted
chiral $N=(4,4)$ multiplet \cite{GHR,IK1}. Thus one can implement various
versions of the partial breaking of $N=(4,4)$ supersymmetry using this
simple system. This extended $D=2$ supersymmetry is related, via 
dimensional reduction, to $N=4~D=3$, $N=2~ D=4$, $N=1~ D=5$  and $N=1~
D=6$ supersymmetries. The relevant $D=2$ Goldstone superfield actions
are expected to describe superstrings and D1-branes associated with these 
supersymmetries and spacetimes. Furthermore, it is of interest to construct
similar PBGS models for heterotic $N=(4,0)$ supersymmetry. 
In both cases, besides the $N=(2,2)$ and $N=(2,0)$ superfield descriptions,
there exists a concise off-shell description in terms of $D=2$ harmonic
superfields \cite{ISu,DKS} manifesting all supersymmetries. Hence, it is
tempting to generalize our construction to harmonic $N=(4,4)$ and $N=(4,0)$
superfields. A natural further step is to try to construct the
$D=2$ PBGS actions with partially broken $N=(8,8)$ and $N=(16,16)$
supersymmetries (and their heterotic truncations). They should be 
relevant to $D=10$ superstrings. 

In the above consideration we systematically identified central 
charges with the generators of shift isometries of the superfields
involved. New possibilities for model-building could come out  
in mixed cases, when some central charges are set to generate 
homogeneous rotational isometries. 

Another line of future study may consist in generalizing  
the Goldstone superfield actions presented in this paper, by adding to 
them covariant couplings of Goldstone superfields to other matter and 
gauge superfields and by coupling them to the appropriate world-sheet 
supergravities. Such extended PBGS models should 
amount to a static-gauge formulation of superstrings and D1-branes evolving
in non-trivial curved backgrounds.       

\vspace{0.5cm}

\noindent{\large\bf  Acknowledgments}\\

The authors are grateful to A.A. Kapustnikov for stimulating
discussions. E.I., S.K. and B.Z. thank the Institut f\"ur Theoretische
Physik, Universit\"at Hannover, for its hospitality. This work was
partially supported by the grants RFBR-DFG-99-02-04022, RFBR-CNRS
98-02-22034, RFBR-99-02-18417, NATO-PST.CLG-974874 and the 
Heisenberg-Landau programme.

%\vfill\eject

\end{document}